\documentclass{tlp}

\usepackage{graphicx}
\usepackage{latexsym}
\usepackage{amssymb}
\usepackage{stmaryrd}
\usepackage{color}
\usepackage{amsmath}

\newcommand{\nop}[1]{}
\newtheorem{definition}{Definition}
\newtheorem{example}{Example}
              
\newtheorem{lemma}{Lemma}               
\newtheorem{theorem}{Theorem}         
\newtheorem{corollary}{Corollary}     
\newtheorem{proposition}{Proposition}

\newcommand{\bldl}{\smallskip\[\tt\begin{array}{ll}}
\newcommand{\cldl}{\vspace{-0.4cm}\[\tt\begin{array}{ll}}
\newcommand{\eldl}{\end{array}\]\rm}

\newcommand{\TT}   {\mbox{$\cal T$}}

\newcommand{\LL}   {\mbox{$\cal L$}}

\newcommand{\Deltac} {\Delta_R}
\def\LL{{\cal L}}
\def\PP{{\cal P}}

\def\MM{{\cal MM}}
\def\SM{{\cal SM}}

\def\wrt{w\mbox{$.$}r\mbox{$.$}t\mbox{$.$}\ }

\def\<{\langle}
\def\>{\rangle}
\def\+{\mbox{+}}
\def\-{\mbox{-}}
\def\={\mbox{\rm =}}
\def\.{\mbox{\rm .}}
\def\({\begin{center}$}
\def\){$\end{center}}
\def\"{``}

\newcommand{\G}{\mbox{${\cal G}$}}

\newcommand{\AR}{\mbox{${\cal AR}$}}   
\newcommand{\FG}{\mbox{${\cal FG}$}}   

\newcommand{\Safe}{\mbox{${\cal SP}$}} 
\newcommand{\AC}{\mbox{${\Gamma\!\cal A}$}}   

\newcommand{\GA}{\mbox{${\Gamma\!A}$}} 
\newcommand{\safe}{\mathit{safe}}
\def\acyclicity{$\Gamma$-acyclicity}
\def\acyclic{$\Gamma$-acyclic}

\submitted{7 December 2012}
\revised{26 March 2014}
\accepted{29 June 2014}

\begin{document}


\title[Checking Termination of the Bottom-Up Evaluation of Logic Programs]{Checking Termination of Bottom-Up Evaluation of Logic Programs with Function
Symbols \footnote{This work refines and extends results from the
conference paper \cite{GrecoST12}.}}

\author[M. Calautti, S. Greco, F. Spezzano and I. Trubitsyna]{
MARCO CALAUTTI, SERGIO GRECO,  \\
{\normalsize \em FRANCESCA SPEZZANO and IRINA TRUBITSYNA}\\
DIMES, Universit\`{a} della Calabria\\
87036 Rende(CS), Italy\\
E-mail: \{calautti,greco,fspezzano,trubitsyna\}@dimes.unical.it
}

\maketitle

\noindent
{\bf Note:} To appear in Theory and Practice of Logic Programming (TPLP).

\label{firstpage}

\begin{abstract}
Recently, there has been an increasing interest in the bottom-up evaluation of
the semantics of logic programs with complex terms.
The presence of function symbols in the program may render
the ground instantiation infinite, and
finiteness of models and termination of the evaluation procedure, in the general case,
are not guaranteed anymore.
Since the program termination problem is undecidable in the general case,
several decidable criteria (called program termination criteria) have been recently proposed.
However, current conditions are not able to identify even simple programs,
whose bottom-up execution always terminates.

The paper introduces new decidable criteria for checking termination of logic
programs with function symbols under bottom-up evaluation, by deeply analyzing the program structure. 
First, we analyze the  propagation of complex terms among
arguments by means of the extended version of the argument graph
called \emph{propagation graph}. The resulting criterion, called
\emph{\acyclicity}, generalizes most of the decidable criteria
proposed so far. Next, we study how rules may activate each other and define a more powerful criterion, called
\emph{safety}. This criterion uses the so-called \emph{safety function} able to  analyze
how rules may activate each other and how the presence of some arguments
in a rule limits its activation.
We also study the application of the proposed criteria to bound
queries and show that the safety criterion is well-suited to identify
relevant classes of programs and bound queries. Finally, we propose
a hierarchy of classes of terminating programs,  called
\emph{$k$-safety}, where the $k$-safe class strictly includes the
$(k\-1)$-safe class.
\end{abstract}



\section{Introduction}
Recently, there has been an increasing interest in the bottom-up evaluation of
the semantics of logic programs with complex terms.
Although logic languages under stable model semantics have enough expressive power to express
problems in the second level of the polynomial hierarchy, in some cases function symbols
make languages compact and more understandable.
For instance, several problems can be naturally expressed using list and set constructors, and arithmetic operators.
The presence of function symbols in the program may render
the ground instantiation infinite, and
finiteness of models and termination of the evaluation procedure, in the general case,
are not guaranteed anymore.
Since the program termination problem is undecidable in the general case,
several decidable sufficient conditions (called \emph{program termination criteria}) have been recently proposed.

The program termination problem has
received a significant attention since the beginning
of logic programming and deductive databases \cite{KrishnamurthyRS96} and
 has recently received an increasing interest.
A considerable body of work has been done on termination of logic programs under top-down evaluation
\cite{Schreye94,Marchiori96,Ohlebusch01,CodishLS05,SerebrenikS05,NguyenGSS07,BruynoogheCGGV07,NishidaV10,Schneider-KampGST09,Schneider-KampGN09,Schneider-KampGSST10,Stroder10,Voets10,BrockschmidtMOG12,LiangK13,Bonatti04,BaseliceBC09}.
In this context, the class of \emph{finitary} programs, allowing decidable (ground)
query computation using a top-down evaluation, has been proposed in \cite{Bonatti04,BaseliceBC09}.
Moreover, there are other research areas, such as these of term rewriting~\cite{Zantema95,SternagelM08,ArtsG00,EndrullisWZ08,FerreiraZ96} and chase termination
\cite{Fagin-TCS,Lausen-09-Err,Marnette-09,GrecoST11,GreSpe10}, whose results can be of interest
to the logic program termination context.

In this paper, we consider logic programs with function symbols \emph{under the stable model semantics}~\cite{GelLif88,GelLif91}
and thus, all the excellent works mentioned above cannot be straightforwardly applied to our setting.
Indeed, the goal of top-down termination analysis is to detect, for a given program and
query goal, sufficient conditions guaranteeing that the resolution algorithm terminates.
On the other side, the aim of the bottom-up termination analysis is to guarantee the existence of
an equivalent finite ground instantiation of the input program.
Furthermore, as stated in \cite{Schreye94}, even restricting our attention to the top-down
approach, the termination of logic programs strictly depends on
the selection and search rules used in the resolution algorithm.
Considering the different aspects of term rewriting and termination of logic programs, we address readers
to \cite{Schreye94} (pages 204-207).

In this framework, the class of \emph{finitely ground  programs} ($\FG$)
has been proposed in \cite{CalimeriCIL08}.
The key property of this class is that stable models (answer sets)  are computable as
for each program $\PP$ in this class, there exists a finite and computable subset 
of its instantiation (grounding), called \emph{intelligent instantiation},
having precisely the same answer sets as $\PP$.
Since the problem of deciding whether a program is in $\FG$ is not decidable,
decidable subclasses, such as
\emph{finite domain programs} \cite{CalimeriCIL08},
\emph{$\omega$-restricted programs} \cite{Syrjanen01},
\emph{$\lambda$-restricted programs} \cite{GebserST07}, and the most general one,
\emph{argument-restricted programs} \cite{LierlerL09}, have been proposed.

Current techniques analyze how values are propagated among predicate
arguments to detect whether a given argument is \emph{limited}, i.e. whether
the set of values which can be associated with the argument, also called \emph{active domain}, is finite.
However, these methods have limited capacity in comprehending that arguments are limited
in the case where different function symbols appear in the recursive rules.
Even the argument-restricted criterion, which is one the most general criteria, fails in such cases.

Thus, we propose a new technique, called \acyclicity , whose aim is to improve the argument-restricted criterion
without changing the (polynomial) time complexity of the argument-restricted criterion.
This technique makes use of the so-called \emph{propagation graph}, that represents the propagation of values
among arguments and the construction of complex terms
during the program evaluation.

Furthermore, since many practical programs are not recognized by current
termination criteria, including the \acyclicity\ criterion,  we propose an even more general technique,
called \emph{safety}, which also analyzes how rules activate each other.
The new technique allows us to recognize as terminating many classical programs, 
still guaranteeing polynomial time complexity.

\begin{example}\label{count-ex}
Consider the following program $P_{\ref{count-ex}}$ computing the length of a list stored in
a fact of the form $\tt input(L)$:
\[
\begin{array}{l}
\tt r_0:\ list(L) \leftarrow input(L). \\
\tt r_1:\ list(L) \leftarrow list([X|L]). \\
\tt r_2:\ count([\,],0). \\
\tt r_3:\ count([X|L], I+1) \leftarrow list([X|L]),\ count(L,I).
\end{array}
\]
where $\tt input$ is a base predicate defined by only one fact of the
form $\tt input([a,b,...])$.~\hfill $\Box$
\end{example}

The safety technique, proposed in this paper, allows us to understand that $P_{\ref{count-ex}}$ is finitely ground
and, therefore, terminating under the bottom-up evaluation.

\paragraph{Contribution}$\!\!$.
\begin{itemize}
\item
We first refine the method proposed in \cite{LierlerL09} by introducing
the set of restricted arguments and we show that the complexity of finding such arguments is polynomial in the size of the given program.
\item We then introduce the class of
\acyclic\ programs, that strictly extends the
class of argument-restricted programs. Its definition is based on a particular graph, called
propagation graph, representing  how complex terms in
non restricted arguments are created and used during the bottom-up
evaluation.
We also show that the complexity of checking whether a program is \acyclic\ is polynomial in the size of the given program.
\item
Next we introduce the \emph{safety function} whose iterative application, starting from
the set of \acyclic\ arguments, allows us to derive a larger set of limited arguments,
by analyzing how rules may be activated.
In particular, we define the \emph{activation graph} that represents
how rules may activate each other and design conditions detecting rules
whose activation cannot cause their head arguments to be non limited.

\item
Since new criteria are defined for normal logic programs without negation,
we extend their application to the case of disjunctive logic programs with negative literals
and show that the computation of stable models can be performed
using current ASP systems, by a simple rewriting of the source program.

\item
We propose the application of the new criteria to bound queries and
show that the safety criterion is well suited to identify relevant classes of programs and bound queries.
\item
As a further improvement,
we introduce the notion of \emph{active paths} of length $k$ and show its applicability
in the termination analysis. In particular, we generalize
the safety criterion and show that the \emph{$k$-safety} criteria
define a hierarchy of terminating criteria for logic programs with function symbols.
\item
Complexity results for the proposed techniques are also presented.
More specifically, we show that the complexity of deciding whether
a program $\PP$ is \acyclic\ or safe is polynomial in the size of $\PP$,
whereas the complexity of the deciding whether a program is $k$-safe, with $k>1$
is exponential.
\end{itemize}
A preliminary version of this paper has been presented at the
28th International Conference on Logic Programming \cite{GrecoST12}.
Although the concepts of \acyclic\ program and safe program have been introduced in the conference paper, the definitions contained in the current version are different. Moreover, most of the theoretical results and all complexity results contained in this paper as well as  the definition of k-safe program are new.

\paragraph{Organization}$\!\!$.
The paper is organized as follows. Section~\ref{sec-preliminaries}
introduces basic notions on logic programming with function symbols.
Section~\ref{section:ar} presents the argument-restriction criterion. 
In Section~\ref{sec:acyclic} the propagation of complex terms among arguments
is investigated and the class of \acyclic\ programs is defined.
Section~\ref{sec:safe} analyzes how rules activate each other and
introduces the \emph{safety} criterion.
In Section~\ref{sec:boundQuery} the applicability of the
safety criterion to (partially) ground queries is discussed.
Section~\ref{sec:k-safe} presents further improvements extending the safety criterion.
Finally, in Section~\ref{sec:compSM} the application of termination criteria to
general disjunctive programs with negated literals is presented.

\section{Logic Programs with Function symbols}\label{sec-preliminaries}

\paragraph{\bf Syntax.}
We assume to have infinite sets of \emph{constants}, \emph{variables}, \emph{predicate symbols}, and \emph{function symbols}.
Each predicate and function symbol $g$ is associated with a fixed \emph{arity}, denoted by $ar(g)$,
which is a non-negative integer for predicate symbols and a natural number for function symbols.

A \emph{term} is either a constant, a variable, or an expression of the form $f(t_1,\dots,t_m)$, where $f$ is a function symbol of arity $m$ and the $t_i$'s are terms.
In the first two cases we say the term is \emph{simple} while in the last case we say it is \emph{complex}.
The binary relation \emph{subterm} over terms is recursively defined as follows:
every term is a subterm of itself; if $t$ is a complex term of the form $f(t_1,\dots,t_m)$, then every $t_i$ is a subterm of $t$ for $1 \leq i \leq m$;
if $t_1$ is a subterm of $t_2$ and $t_2$ is a subterm of $t_3$, then $t_1$ is a subterm of $t_3$.
The depth $d(u,t)$ of a simple term $u$ in a term $t$ that contains $u$ is recursively defined as follows:
\[\begin{array}{l}
d(u,u) = 0,\\
d(u,f(t_1,\.\.\.,t_m)) = 1 + \max\limits_{{i\ :\ t_i\, contains\, u}} d(u,t_i).
\end{array}
\]
The \emph{depth of term} $t$, denoted by $d(t)$, is the maximal depth of all simple terms occurring in $t$.

An \emph{atom} is of the form $p(t_1,\dots,t_n)$, where $p$ is a predicate symbol of arity $n$ and the $t_i$'s are terms (we also say that the atom is a $p$-atom).
A \emph{literal} is either an atom $A$ (\emph{positive} literal) or its negation $\neg A$ (\emph{negative} literal).

A \emph{rule} $r$ is of the form:
\(
A_1 \vee \.\.\. \vee A_m \leftarrow B_1,\.\.\., B_k, \neg
C_1,\.\.\., \neg C_n
\)
\noindent
where $m >0$,  $k\geq 0$, $n \geq 0$, and $A_1,\.\.\. ,A_m,B_1,\.\.\.,B_k,$ $C_1,$ $\.\.\.,C_n$ are
atoms. The disjunction $A_1 \vee \.\.\. \vee A_m$ is called the
\emph{head} of $r$ and is denoted by $head(r)$; the
conjunction $B_1,\.\.\., B_k, \neg C_1,\.\.\., \neg C_n$ is
called the \emph{body} of $r$ and is denoted by $body(r)$.
The \emph{positive} (resp. \emph{negative}) \emph{body} of $r$ is the conjunction
$B_1,\.\.\., B_k$ (resp. $\neg C_1,\.\.\., \neg C_n$) and is denoted by
$body^{+}(r)$ (resp. $body^{-}(r)$).
With a slight abuse of notation we use $head(r)$ (resp. $body(r)$, $body^+(r)$, $body^-(r)$) to also denote the \emph{set} of atoms (resp. literals) appearing in the head (resp. body, positive body, negative body) of $r$.
If $m=1$, then $r$ is {\em normal}; if $n=0$, then $r$ is {\em positive}. If a rule $r$ is both normal and positive, then it is \emph{standard}.

A \emph{program} is a finite set of rules.
A program is \emph{normal} (resp. \emph{positive}, \emph{standard}) if every rule in it is normal (resp. positive, standard).
A term (resp. an atom, a literal, a rule, a program) is said to be {\em ground} if no variables occur in it.
A ground normal rule with an empty body is also called a \emph{fact}.
For any atom $A$ (resp. set of atoms, rule), $var(A)$ denotes the set of variables occurring in $A$.

We assume that programs are \emph{range restricted}, i.e., for each rule, the variables
appearing in the head or in negative body literals also appear in some positive body literal.

The \emph{definition} of a predicate symbol $p$ in a program $\PP$ consists of all rules in $\PP$ with $p$ in the head.
Predicate symbols are partitioned into two different classes: \emph{base} predicate symbols, whose definition can
contain only facts (called \emph{database facts}), and \emph{derived} predicate symbols, whose definition can contain any rule.
Database facts are not shown in our examples as they are not relevant for the proposed criteria.

Given a program $\PP$, a predicate $p$ depends on a predicate $q$
if there is a rule $r$ in $\PP$ such that $p$ appears in the head and
$q$ in the body, or there is a predicate $s$ such that $p$
depends on $s$ and $s$ depends on $q$. A predicate $p$ is said
to be \emph{recursive} if it depends on itself, whereas two predicates
$p$ and $q$ are said to be \emph{mutually recursive} if $p$ depends on
$q$ and $q$ depends on $p$.
A rule $r$ is said to be \emph{recursive} if its body contains a predicate symbol mutually recursive with a predicate symbol in the head.
Given a rule $r$, $rbody(r)$ denotes the set of body atoms whose predicate symbols are mutually recursive with
the predicate symbol of an atom in the head.
We say that $r$ is \emph{linear} if $|rbody(r)| \leq 1$.
We say that a recursive rule $r$ defining a predicate $p$ is \emph{strongly} linear if
it is linear, the recursive predicate symbol appearing in the body is $p$
and there are no other recursive rules defining $p$.
A predicate symbol $p$ is said to be linear (resp. strongly linear)
if all recursive rules defining $p$ are linear (resp. strongly linear).

A {\em substitution} is a finite set of pairs  $\theta = \{
X_1/t_1,\.\.\.,X_n/t_n \}$ where $t_1,\.\.\.,t_n$ are terms
and $X_1,\.\.\.,X_n$ are distinct variables not occurring in $t_1,\dots,t_n$.
If $\theta = \{ X_1/t_1,\.\.\.,X_n/t_n \}$ is a substitution and $T$ is a term
or an atom, then $T \theta$ is the term or atom obtained from
$T$ by simultaneously replacing each occurrence of $X_i$ in
$T$ by $t_i$ ($1 \leq i \leq n$) --- $T\theta$ is called an
{\em instance} of $T$.
Given a set $S$ of terms (or atoms), then $S\theta=\{T\theta \mid T \in S\}$.
A substitution $\theta$ is a {\em unifier} for a finite set of terms (or atoms)
$S$ if $S\theta$ is a  singleton.
We say that a set of terms (or atoms) $S$ {\em unify} if there exists a unifier
$\theta$ for $S$.
Given two substitutions $\theta=\{X_1/t_1,\dots,X_n/t_n\}$ and $\vartheta=\{Y_1/u_1,\dots,Y_m/u_m\}$, their \emph{composition}, denoted $\theta\circ\vartheta$, is the substitution obtained from the set $\{X_1/t_1\vartheta,\dots,X_n/t_n\vartheta,$ $Y_1/u_1,\dots,Y_m/u_m\}$ by removing every $X_i/t_i\vartheta$ such that $X_i=t_i\vartheta$ and every $Y_j/u_j$ such that $Y_j \in \{X_1,\dots,X_n\}$.
A substitution $\theta$ is \emph{more general} than a substitution $\vartheta$ if there exists a substitution $\eta$ such that $\vartheta=\theta\circ\eta$.
A unifier $\theta$ for a set $S$ of terms (or atoms) is called a {\em
most general unifier} (mgu) for $S$ if it is more general than any other unifier for $S$.
The mgu is unique modulo renaming of variables.

\paragraph{\bf Semantics.}
Let $\PP$ be a program.
The \emph{Herbrand universe} $H_{\cal P}$ of $\PP$ is the
possibly infinite set of ground terms which can be built using
constants and function symbols appearing in $\PP$. The \emph{Herbrand base}
$B_{\cal P}$ of $\PP$ is the set of ground atoms which can
be built using predicate symbols appearing in $\PP$ and ground terms
of $H_{\cal P}$.
A rule $r'$ is a {\em ground instance} of a rule $r$ in $\PP$ if $r'$ can be obtained from $r$ by substituting every variable in $r$ with some ground term in $H_{\mathcal P}$.
We use $ground(r)$ to denote the set of all ground instances of $r$ and  $ground(\PP)$ to denote the set of all ground instances of the rules in $\PP$, i.e., $ground(\PP)=\cup_{r\in\PP} ground(r)$.
An \emph{interpretation} of $\PP$ is any subset $I$ of $B_{\cal P}$.
The truth value of a ground atom $A$ \wrt $I$, denoted $value_I(A)$, is \emph{true} if $A \in I$, \emph{false} otherwise.
The truth value of $\neg A$ \wrt $I$, denoted $value_I(\neg A)$, is \emph{true} if $A \not\in I$, \emph{false} otherwise.
The truth value of a conjunction of ground literals $C = L_1,\.\.\.,L_n$ \wrt $I$ is $value_I(C) = \min(\{value_I(L_i)\ |\ 1 \leq i \leq n\})$---here the ordering \emph{false} $<$ \emph{true} holds---whereas the truth value of a disjunction of ground literals $D = L_1 \vee \.\.\. \vee L_n$ \wrt $I$ is $value_I(D)\! =\! \max(\{value_I(L_i)\ |\ 1 \leq i \leq n \})$; if $n=0$, then $value_I(C)\! =\!$ \emph{true} and $value_I(D)\!=$ \emph{false}.
A ground rule $r$ is {\em satisfied} by $I$, denoted $I \models r$, if $value_I(head(r)) \geq value_I(body(r))$;
we write $I\not\models r$ if $r$ is not satisfied by $I$.
Thus, a ground rule $r$ with empty body is satisfied by $I$ if $value_I(head(r))\! = $ \emph{true}.
An interpretation of $\PP$ is a \emph{model} of $\PP$ if it satisfies every ground rule in $ground(\PP)$.
A model $M$ of $\PP$ is minimal if no proper subset of $M$ is
a model of $\PP$.
The set of minimal models of $\PP$ is denoted by $\MM(\PP)$.

Given an interpretation $I$ of $\PP$, let $\PP^I$ denote the ground positive program derived
from $ground(\PP)$ by \emph{(i)} removing every rule containing a
negative literal $\neg A$ in the body with $A \in I$, and \emph{(ii)}
removing all negative literals from the remaining rules.
An interpretation $I$ is a \emph{stable model} of $\PP$ if
and only if $I \in \MM(\PP^I)$ \cite{GelLif88,GelLif91}.
The set of stable models of $\PP$ is denoted by $\SM(\PP)$.
It is well known that stable
models are minimal models (i.e., $\SM(\PP) \subseteq \MM(\PP)$).
Furthermore, minimal and stable model semantics coincide for positive programs  (i.e., $\SM(\PP) = \MM(\PP)$).
A standard program has a unique minimal model, called \emph{minimum model}.

 Given a set of ground atoms $S$ and a predicate $g$ (resp. an atom
$A$), $S[g]$ (resp. $S[A]$) denotes the set of $g$-atoms
(resp. ground atoms unifying with $A$) in $S$. Analogously, for a given
set $M$ of sets of ground atoms, we shall use the following
notations $M[g] = \{ S[g]\ |\ S \in M \}$ and $M[A] = \{
S[A]\ |\ S \in M \}$. Given a set of ground atoms $S$, and a set $G$ of predicates symbols, then $S[G] = \cup_{g \in G} S[g]$.

\paragraph{\bf Argument graph. }

Given an $n$-ary predicate $p$, $p[i]$ denotes the $i$-th argument of $p$, for $1 \leq i \leq n$.
If $p$ is a base (resp. derived) predicate symbol, then $p[i]$ is said to be a \emph{base} (resp. \emph{derived}) argument.
The set of all arguments of a program $\PP$ is denoted by $args(\PP)$;
analogously, $args_b(\PP)$ and $args_d(\PP)$ denote the sets of all base and derived arguments, respectively.

For any program $\PP$ and n-ary predicate $p$ occurring in $\PP$,
an argument $p[i]$, with $1 \leq i \leq n$, is associated with the set of values it can take
during the evaluation;
this domain, called \emph{active domain} of $p[i]$,
is denoted by $AD(p[i]) = \{ t_i | p(t_1,\dots,t_n) \in M \wedge M \in \SM(\PP) \}$.
An argument $p[i]$ is said to be \emph{limited} iff $AD(p[i])$ is finite.

The \emph{argument graph} of a program $\PP$, denoted $G(\PP)$, is a directed graph whose nodes are $args(\PP)$ (i.e. the arguments of $\PP$),
and there is an edge from $q[j]$ to $p[i]$, denoted by $(q[j],p[i])$, iff there is a rule $r \in \PP$ such that:
\begin{enumerate}
\vspace*{-2mm}
\item
an atom $p(t_1,\.\.\.,t_n)$ appears in $head(r)$,
\item
an atom $q(u_1,\.\.\.,u_m)$ appears in $body^{+}(r)$, and
\item
terms $t_i$ and $u_j$ have a common variable.
\end{enumerate}
Consider, for instance, program $P_{\ref{count-ex}}$ of Example \ref{count-ex}.
$G(P_{\ref{count-ex}}) = (args(P_{\ref{count-ex}}),E)$, where
$args(P_{\ref{count-ex}}) = \tt \{ input[1], list[1], count[1], count[2] \}$, whereas,
considering the occurrences of variables in the rules of $P_{\ref{count-ex}}$ we have that
$E =$
$\tt \{ (input[1],list[1]),\ (list[1],list[1]),\ (list[1],count[1]),\ (count[1],count[1]),\ \\(count[2],count[2]) \}$.

\paragraph{\bf Labeled directed graphs.}
In the following we will also consider labeled directed graphs, i.e. directed graphs with
labeled edges. In this case we represent an edge from $a$ to $b$ as a triple
$(a,b,l)$, where $l$ denotes the label.

A \emph{path} $\pi$ from $a_1$ to $b_m$ in a possibly labeled directed
graph is a non-empty sequence
$(a_1,b_1,l_1), \dots,$ $(a_m,b_m,
l_m)$ of its edges
s.t. $b_i=a_{i+1}$ for all $1 \leq i < m$; if the first and
last nodes coincide (i.e., $a_1=b_m$), then $\pi$ is called a
\emph{cyclic path}.
In the case where the indication of the starting edge is not
relevant, we will call a cyclic path a \emph{cycle}.

We say that a node $a$  \emph{depends on} a
node $b$ in a graph iff there is a path from $b$ to
$a$ in that graph.
Moreover, we say that $a$ \emph{depends on} a cycle $\pi$ iff it depends on a node $b$ appearing in $\pi$.
Clearly, nodes belonging to a cycle $\pi$ depend on $\pi$.

\section{Argument ranking}\label{section:ar}

The argument ranking of a program has
been proposed in \cite{LierlerL09} to define the
class $\AR$ of \emph{argument-restricted} programs.

An \emph{argument ranking} for a program $\PP$ is a partial function $\phi$ from $args(\PP)$ to non-negative integers, called \emph{ranks}, such that, for every rule $r$ of $\PP$, every atom $p(t_1,\dots,t_n)$ occurring in the head of $r$, and every variable $X$ occurring in a term $t_i$, if $\phi(p[i])$ is defined, then $body^{+}(r)$ contains an atom
$q(u_1,\dots,u_m)$ such that $X$ occurs in a term $u_j$, $\phi(q[j])$ is defined, and the following condition is satisfied
\begin{equation}\label{eq:ar}
 \phi(p[i]) - \phi(q[j]) \geq d(X,t_i) - d(X,u_j).
\end{equation}
 A program $\PP$ is said to be \emph{argument-restricted} if
 it has an argument ranking  assigning ranks to all arguments of $\PP$.

\begin{example}\label{Example-AR1}
Consider the following program $P_{\ref{Example-AR1}}$, where $\tt b$ is a base predicate:
\[
\begin{array}{l}
r_1: \tt p(f(X)) \leftarrow p(X), b(X)\. \\
r_2: \tt t(f(X)) \leftarrow p(X)\. \\
r_3: \tt s(X)    \leftarrow t(f(X))\.\\
\end{array}
\]
This program has an argument ranking $\phi$, where
$\phi(\tt b[1])$$=0$, $\phi(\tt p[1])$$=1$, $\phi(\tt t[1])$$=2$ and $\phi(\tt s[1])$$=1$.
Consequently, $P_{\ref{Example-AR1}}$  is argument-restricted.~\hfill~$\Box$
\end{example}

Intuitively, the rank of an argument is an estimation
of the depth of terms that may occur in it. 
In particular, let $d_1$ be the rank assigned to a given argument $p[i]$
and let $d_2$ be the maximal depth of terms occurring in the database facts.
Then $d_1+d_2$ gives an upper bound of the depth of terms that may occur in
$p[i]$ during the program evaluation.
Different argument rankings may satisfy condition~(\ref{eq:ar}).
A function assigning minimum ranks to arguments is denoted by $\phi_{min}$.

\paragraph{\bf Minimum ranking.}
We define a monotone operator $\Omega$ that takes as input a function $\phi$ over arguments and gives as output a function over arguments that gives an upper bound of the depth of terms. \\
More specifically, we define $\Omega(\phi)(p[i])$ as
\[
max ( max \{ D(p(t_1,\dots,t_n),r,i,X)\, |\, r \in \PP \wedge p(t_1,\dots,t_n) \in head(r) \wedge X\,occurs\,in\,t_i  \}, 0)
\]
where $D(p(t_1,\dots,t_n),r,i,X)$ is defined as
\[
min \{ d(X,t_i) - d(X,u_j) + \phi(q[j])\,|\, q(u_1,\dots,u_m) \in body^+(r) \wedge\,  X\,occurs\,in\,u_j \}.
\]

In order to compute $\phi_{min}$
we compute the fixpoint of $\Omega$ starting from the
function $\phi_0$ that assigns $0$ to all arguments.
In particular, we have:
\[
\begin{array}{l}
\phi_0(p[i]) = 0; \\ 
\phi_{k}(p[i]) = \Omega(\phi_{k-1})(p[i]) = \Omega^k(\phi_0)(p[i]).
\end{array}
\]

The function $\phi_{min}$ is defined as follows:

{\small
\[
\phi_{min}(p[i]) =
\left\{
   \begin{array}{ll}
      \Omega^k(\phi_0)(p[i]) \ \ \ & if\ \exists k \mbox{ (finite) s.t. } \Omega^k(\phi_0)(p[i]) = \Omega^{\infty}(\phi_0) (p[i]) \\
      undefined & otherwise
   \end{array}
\right.
\]
}

We denote the set of \emph{restricted arguments} of $\PP$ as $AR(\PP)=\{p[i]\ | \ p[i] \in args(\PP) \wedge \ \phi_{min}(p[i]) \mbox{ is defined}\}$.
Clearly, from definition of $\phi_{min}$, it follows that all restricted arguments are limited.
Observe that $\PP$ is \emph{argument-restricted} iff $AR(\PP) = args(\PP)$.

\begin{example}\label{Example-AR2}
Consider again program $P_{\ref{Example-AR1}}$ from Example~\ref{Example-AR1}. The following table shows the
first four iterations of $\Omega$ starting from the base ranking function $\phi_0$:
\begin{table}[ht]
\centering 
\begin{tabular}{c c c c c c } 
\hline 
       & \hspace*{7mm}$\phi_0$\hspace*{7mm} & $\phi_1=\Omega(\phi_0)$ & $\phi_2=\Omega(\phi_1)$ & $\phi_3=\Omega(\phi_2)$ & $\phi_4=\Omega(\phi_3)$  \\
\hline 
$\tt b[1]$ &  0 & 0 & 0 & 0 & 0 \\ 
$\tt p[1]$ & 0 & 1 & 1 & 1 & 1 \\
$\tt t[1]$ & 0 & 1 & 2 & 2 & 2 \\
$\tt s[1]$ & 0 & 0 & 0 & 1 & 1 \\
\hline 
\end{tabular}
\label{table:nonlin} 
\end{table}

\noindent
Since $\Omega(\phi_3)= \Omega(\phi_2)$, further applications of $\Omega$ provide the same result. Consequently, $\phi_{min}$ coincides with $\phi_3$ and defines ranks for all arguments of $P_{\ref{Example-AR1}}$.~\hfill~$\Box$
\end{example}

Let $M = |args(\PP)| \times d_{max}$, where $d_{max}$ is the largest depth of terms occurring in the heads of rules of $\PP$.
One can determine whether $\PP$ is argument-restricted by
iterating $\Omega$ starting from $\phi_0$ until
\begin{description}
   \item[-] one of the values of $\Omega^k(\phi_0)$ exceeds $M$, in such a case $\PP$ is not argument-restricted;
   \item[-] $\Omega^{k+1}(\phi_0) =\Omega^k(\phi_0)$, in such a case $\phi_{min}$ coincides with $\phi_k$, $\phi_{min}$ is total, and $\PP$ is argument-restricted.
\end{description}
Observe that if the program is not argument-restricted the first condition is verified with $k\leq M\times |args(\PP)|\leq M^2$, as at each iteration the value assigned to at least one argument is changed.
Thus, the problem of deciding whether a given program $\PP$ is argument-restricted is in $PTime$.
In the following section we will show that the computation of restricted arguments can be done in polynomial time
also when $\PP$ is not argument-restricted (see Proposition~\ref{prop:numberIter}).

\section{\acyclic\ programs}\label{sec:acyclic}

In this section we exploit the
role of function symbols for checking program termination under bottom-up
evaluation. Starting from this section, we will consider standard logic programs.
Only in Section~\ref{sec:compSM} we will refer to general programs,
as it discusses how termination criteria defined for standard programs
can be applied to general disjunctive logic programs with negative literals.
We also assume that if the same variable $X$ appears in two
terms occurring in the head and body of a rule respectively, then at most one of the two
terms is a complex term and that the nesting level of
complex terms is at most one.
As we will see in Section~\ref{sec:compSM}, there is no real restriction in such an assumption as every program
could be rewritten into an equivalent program satisfying such a condition.

The following example shows a program admitting a finite minimum model,
but the argument-restricted criterion is not able to detect it. Intuitively, the definition of argument restricted programs does not take into account the possible presence of different function symbols in the program that may prohibit the propagation of values in some rules and, consequently, guarantee the termination of the bottom-up computation.

\begin{example}\label{Example-notAmmisLabel1}
Consider the following program $P_{\ref{Example-notAmmisLabel1}}$:
\[
\begin{array}{l}
r_0: \tt s(X) \leftarrow b(X)\. \\
r_1: \tt r(f(X)) \leftarrow s(X)\. \\
r_2: \tt q(f(X)) \leftarrow r(X)\. \\
r_3: \tt s(X)    \leftarrow q(g(X))\.
\end{array}
\]
where $\tt b$ is a base predicate symbol.
The program is not argument-restricted since the argument ranking function $\phi_{min}$
cannot assign any value to $\tt r[1]$, $\tt q[1]$, and $\tt s[1]$.
However the bottom-up computation always terminates, independently from the database instance.
~\hfill~$\Box$
\end{example}

In order to represent the propagation of values among arguments, we introduce the concept
of \emph{labeled argument graphs}.
Intuitively, it is an extension of the argument graph where each edge has a label describing
how the term propagated from one argument to another changes.
Arguments that are not dependent on a cycle can propagate a finite number of values ​​and,
therefore, are limited.

Since the active domain of limited arguments is finite, we can delete edges ending in the corresponding nodes from the
labeled argument graph.
Then, the resulting graph, called \emph{propagation graph}, is deeply analyzed to identify
further limited arguments.

\begin{definition}[Labeled argument graph]\label{agument-graph1}
Let $\PP$ be a program. The \emph{labeled argument graph}
$\G_L(\PP)$ is a labeled directed graph $(args(\PP),E)$
where $E$ is a set of labeled edges defined as follows.
For each pair of nodes $p[i], q[j] \in args(\PP)$ such that there is a rule $r$ with
$head(r) = p(v_1,\dots,v_n)$,
$q(u_1,\dots,u_m) \in body(r)$, and
terms $u_j$ and $v_i$ have a common variable $X$,
there is an edge $(q[j],p[i],\alpha) \in E$ such that
\begin{itemize}
\vspace*{-1mm}
\item
$\alpha = \epsilon$ \ if $u_j = v_i = X$,  
\item
$\alpha = f$ \ if $u_j = X$ and $v_i = f(\.\.\.,X,\.\.\.)$,
\item
$\alpha = \overline{f}$ \ if $u_j = f(\.\.\.,X,\.\.\.)$ and $v_i = X$.
\hfill $\Box$
\end{itemize}
\end{definition}

In the definition above, the symbol $\epsilon$ denotes the empty label which concatenated to a string does not modify the
string itself, that is, for any string $s$, $s \epsilon = \epsilon s = s$.

\begin{figure}
  \includegraphics[width=12cm]{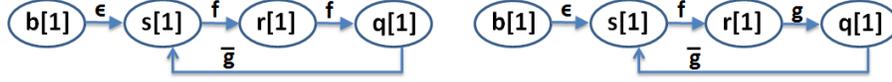}\\
  \caption{Labeled argument graphs of programs $P_{\ref{Example-notAmmisLabel1}}$ (left) and  $P_{\ref{ex:label-path}}$ (right)}\label{FigureEs4}
\end{figure}

The labeled argument graph of program $P_{\ref{Example-notAmmisLabel1}}$
is shown in Figure~\ref{FigureEs4}~(left). 
The edges of this graph represent how the propagation of values occurs.
For instance, edge $\tt (b[1], s[1], \epsilon)$ states that a term $\tt t$ is propagated without
changes from $\tt b[1]$ to $\tt s[1]$ if rule $r_0$ is applied;
analogously, edge $\tt (s[1], r[1], f)$ states that starting from a term $\tt t$ in $\tt s[1]$ we obtain $\tt f(t)$
in $\tt r[1]$ if rule $r_1$ is applied,
whereas edge $\tt (q[1],s[1], \overline{g})$ states that starting from a term $\tt g(t)$ in $\tt q[1]$ we obtain $\tt t$
in $\tt s[1]$ if rule $r_3$ is applied.

\vspace*{2mm}
Given a path $\pi$ in $\G_L(\PP)$ of the form  $(a_1,b_1,\alpha_1), \dots,$
$(a_m,b_m, \alpha_m)$, we denote with $\lambda(\pi)$ the string $\alpha_1 \, \.\.\. \, \alpha_m$.
We say that \emph{$\pi$ spells a
string $w$} if $\lambda(\pi) = w$.
Intuitively, the string $\lambda(\pi)$ describes a sequence of function symbols used to compose and decompose complex terms during the propagation of values among the arguments in $\pi$.

\begin{example}\label{ex:label-path}
Consider program  $P_{\ref{ex:label-path}}$ derived from program $P_{\ref{Example-notAmmisLabel1}}$
of Example \ref{Example-notAmmisLabel1} by replacing rule $r_2$ with the rule $\tt q(g(X)) \leftarrow r(X)$.
The labeled argument graph $\G_L(P_{\ref{ex:label-path}})$ is reported in Figure~\ref{FigureEs4}~(right).
Considering the cyclic path $\pi=\tt (s[1],r[1],f),$ $\tt (r[1],q[1],g),$ $\tt (q[1],s[1],\overline{g})$,
$\lambda(\pi) = \tt fg\overline{g}$ represents  the fact that starting from a term $\tt t$ in $\tt s[1]$ we may obtain the term $\tt f(t)$ in $\tt r[1]$, then we may obtain term $\tt g(f(t))$ in $\tt q[1]$, and term $\tt f(t)$ in $\tt s[1]$, and so on.
Since we may obtain a larger term in $\tt s[1]$, the arguments depending on this cyclic path may not be limited.

Consider now  program $P_{\ref{Example-notAmmisLabel1}}$, whose labeled argument graph is
shown in Figure~\ref{FigureEs4}~(left),  and
the cyclic path $\pi'=\tt (s[1],r[1],f),$ $\tt (r[1],q[1],f),$ $\tt (q[1],s[1],\overline{g})$.
Observe that starting from a term $\tt t$ in $\tt s[1]$ we may obtain term $\tt f(t)$ in $\tt r[1]$ (rule $r_1$), then we may obtain term $\tt f(f(t))$ in $\tt q[1]$ (rule $r_2$). At this point the propagation in this cyclic path terminates since the head atom of rule $r_2$ containing term $\tt f(X)$ cannot match with the body atom of rule $r_3$ containing term $\tt g(X)$.
The string $\lambda(\pi') = \tt ff\overline{g}$ represents the propagation described above.
Observe that for this program all arguments are limited.
~\hfill $\Box$
\end{example}

Let $\pi$ be  a path from $p[i]$ to $q[j]$ in the labeled argument graph.
Let $\hat{\lambda}(\pi)$ be the string
obtained from $\lambda(\pi)$ by iteratively eliminating
pairs of the form $\alpha\overline{\alpha}$  until the resulting string cannot be further reduced.
If $\hat{\lambda}(\pi)=\epsilon$, then starting from a term $t$ in $p[i]$ we obtain the same term $t$ in $q[j]$.
Consequently, if $\hat{\lambda}(\pi)$
is a non-empty sequence of  function symbols $f_{i_1},f_{i_2}\dots, f_{i_k}$, then starting from a term $t$ in $p[i]$ we may obtain a larger term in $q[j]$.
For instance, if $k=2$ and $f_{i_1}$ and $f_{i_2}$ are of arity one, we may obtain $f_{i_2}(f_{i_1}(t))$ in $q[j]$.
Based on this intuition we introduce now a grammar $\Gamma_{\!\cal P}$ in order to distinguish the sequences of function symbols used to compose and decompose complex terms in a program $\PP$,
such that starting from a given term we obtain a larger term.

\vspace*{2mm}
Given a program $\PP$, we denote with $F_{\! \cal P} = \{ f_1,\.\.\.,f_m\}$
the set of function symbols occurring in $\PP$, whereas
$\overline{F}_{\! \cal P} = \{ \overline{f}\ |\ f \in F_{\!\cal P} \}$ and $T_{\! \cal P} = F_{\! \cal P} \cup \overline{F}_{\! \cal P}$.

\begin{definition}\label{def-language}
Let $\PP$ be a program, the  \emph{grammar} $\Gamma_{\!\cal P}$ is a 4-tuple  $(N,T_{\!\cal P},R,S)$,
where
$N = \{ S, S_1, S_2 \}$ is the set of nonterminal symbols,
$S$ is the start symbol, and
$R$ is the set of production rules defined below:
\begin{enumerate}
\item
$S \ \rightarrow S_1 \, f_i \, S_2$, \hspace*{15mm} $\forall f_i \in F_{\!\cal P}$;
\item
$S_1 \rightarrow f_i\,  S_1\, \overline{f}_i\, S_1\ | \ \epsilon$, \hspace*{6,2mm} $\forall f_i \in F_{\!\cal P}$;
\item
$S_2 \rightarrow S_1 \, S_2 \ | \ f_i\, S_2 \ | \ \epsilon$, \hspace*{3mm} $\forall f_i \in F_{\!\cal P}$.
\hfill $\Box$
\end{enumerate}
\end{definition}

The language $\LL(\Gamma_{\!\cal P})$ is the set of strings generated by $\Gamma_{\!\cal P}$.

\begin{example}
Let $F_\PP=\tt \{f, g, h\}$ be the set of function symbols occurring in a program $\PP$.
Then strings $\tt f$, $\tt fg\overline{g}$, $\tt g\overline{g}f$, $\tt fg\overline{g}h\overline{h}$, $\tt fhg\overline{g}\overline{h}$ belong to $\LL(\Gamma_{\!\cal P})$ and represent,
assuming that $\tt f$ is a unary function symbol, different ways to obtain term $\tt f(t)$ starting from term~$\tt t$.~\hfill $\Box$
\end{example}

Note that only if a path $\pi$ spells a string $w \in \LL(\Gamma_{\!\cal P})$,
then starting from a given term
in the first node of $\pi$ we may obtain a larger term in the last node of $\pi$. Moreover, if this path is cyclic, then the arguments depending on it may not be limited. On the other hand, all arguments not depending on a cyclic path $\pi$ spelling
a string $w \in \LL(\Gamma_{\!\cal P})$ are limited.

Given a program $\PP$ and a set of arguments $\cal S$ recognized as limited by a specific criterion, 
the \emph{propagation graph} of $\PP$ with respect to $\cal S$, denoted by $\Delta(\PP,{\cal S})$, consists
of the subgraph derived from $\G_L(\PP)$ by deleting edges ending in a node of $\cal S$.
Although we can consider any set $\cal S$ of limited arguments,
in the following we assume ${\cal S} = AR(\PP)$ and, for the simplicity of notation, we denote
$\Delta(\PP,AR(\PP))$ as $\Delta(\PP)$.
Even if more general termination criteria have been defined in the literature, here we
consider the $AR$ criterion since it is the most general among those so far proposed having polynomial time complexity.

\begin{definition}[\acyclic \ Arguments and \acyclic \ Programs]\label{def-acyclic}
Given a program $\PP$, the set of its \emph{\acyclic\ arguments}, denoted by $\GA(\PP)$,
consists of  all arguments of $\PP$ not depending on a cyclic path in $\Delta(\PP)$
spelling a string of $\LL(\Gamma_{\!\cal P})$.
A program $\PP$ is called \emph{\acyclic} if $\GA(\PP) = args(\PP)$, i.e. if there is no cyclic
path in $\Delta(\PP)$ spelling a string of $\LL(\Gamma_{\!\cal P})$.
We denote the class of \acyclic\  programs  $\AC$.~\hfill~$\Box$
\end{definition}

Clearly, $AR(\PP)  \subseteq \GA(\PP)$, i.e. the set of restricted arguments is contained in the set of \acyclic\ arguments.
As a consequence, the set of argument-restricted programs is a subset of the set of \acyclic\  programs.
Moreover, the  containment is strict, as there exist programs that are \acyclic, but not argument-restricted.
For instance, program $P_{\ref{Example-notAmmisLabel1}}$ from Example \ref{Example-notAmmisLabel1}
is \acyclic, but not argument-restricted.
Indeed, all cyclic paths in $\Delta(P_{\ref{Example-notAmmisLabel1}})$ do not spell strings belonging to the language
$\LL(\Gamma_{P_{\ref{Example-notAmmisLabel1}}})$.

\vspace*{2mm}
The importance of considering the propagation graph instead
of the labeled argument graph in Definition~\ref{def-acyclic}
is shown  in the following example.

\begin{figure}
  \includegraphics[width=11cm]{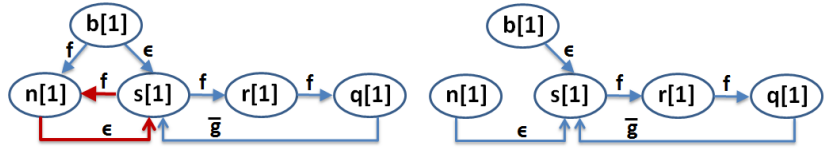}\\
  \caption{Labeled argument graph (left) and propagation graph (right)  of program $P_{\ref{Example-safe-AR}}$ } \label{FigureEs9}
\end{figure}
\newpage
\begin{example}\label{Example-safe-AR}
Consider program $P_{\ref{Example-safe-AR}}$ below
obtained from $P_{\ref{Example-notAmmisLabel1}}$ by adding rules $r_4$ and $r_5$.
\[
\begin{array}{l}
r_0: \tt s(X) \leftarrow b(X)\. \\
r_1: \tt r(f(X)) \leftarrow s(X)\. \\
r_2: \tt q(f(X)) \leftarrow r(X)\. \\
r_3: \tt s(X)    \leftarrow q(g(X))\.\\
r_4: \tt n(f(X))    \leftarrow s(X),b(X)\. \\
r_5: \tt s(X)    \leftarrow n(X)\.
\end{array}
\]
The corresponding labeled argument graph $\G_L(P_{\ref{Example-safe-AR}})$
and propagation graph $\Delta(P_{\ref{Example-safe-AR}})$
are reported  in Figure~\ref{FigureEs9}.
Observe that arguments $\tt n[1]$ and $\tt s[1]$ are involved in the red cycle in the
labeled argument graph $\G_L(P_{\ref{Example-safe-AR}})$
spelling a string of $\LL(\Gamma_{\! P_{\ref{Example-safe-AR}}})$.
At the same time this cycle is not present in the propagation graph
$\Delta(P_{\ref{Example-safe-AR}})$ since
$AR(P_{\ref{Example-safe-AR}})=\tt \{b[1],n[1]\}$
and the program is \acyclic.~\hfill~$\Box$
\end{example}

\begin{theorem}\label{acyclic-finite-domain}
Given a program $\PP$,
\begin{enumerate}\setlength{\itemsep}{0pt}
\vspace*{-2mm}
\item all arguments in $\GA(\PP)$ are limited;
\item if $\PP$ is \acyclic, then $\PP$ is finitely ground. 
\end{enumerate}
\end{theorem}

\vspace*{-3mm}
\proof{
\emph{1)}
As previously recalled, arguments in $AR(\PP)$ are limited.
Let us now show that all arguments in  $\GA(\PP)\setminus AR(\PP)$ are limited too.
Suppose by contradiction that $q[k]\in \GA(\PP)\setminus AR(\PP)$ is not limited.
Observe that depth of terms that may occur in $q[k]$ depends
on the paths in the propagation graph $\Delta(\PP)$ that ends in $q[k]$. In particular, this depth may be infinite only if there is a path $\pi$ from an argument $p[i]$ to $q[k]$ (not necessarily distinct from $p[i]$), such that $\hat{\lambda}(\pi)$ is a string of an infinite length composed by symbols in $F_{P}$. But this is possible only if this path contains a cycle spelling a string in $\LL(\Gamma_{\!\cal P})$. Thus we obtain contradiction with Definition~\ref{def-acyclic}.\\
\emph{2)}
From the previous proof, it follows that every argument in the \acyclic\ program  can take values only from a finite domain. Consequently,
the set of all possible ground terms derived during the grounding process is finite and every \acyclic\ program is finitely ground.
~\hfill $\Box$
}
\\

From the previous theorem we can also conclude that all \acyclic\  programs admit a finite minimum model, as this is a property of finitely ground programs.

\vspace*{2mm}
We conclude by observing that since the language $\LL(\Gamma_{\!\cal P})$ is context-free,
the analysis of paths spelling strings in $\LL(\Gamma_{\!\cal P})$ can be carried out using
pushdown automata. \\

As $\Gamma_{\!\cal P}$ is context
free, the language $\LL(\Gamma_{\!\cal P})$ can be recognized by
means of a pushdown automaton $M = (\{q_0, q_F\}, T_{\!\cal P}, \Lambda, \delta, q_0, Z_0, \{ q_F \} \})$,
where $q_0$ is the initial state, $q_F$ is the final state,
$\Lambda = \{ Z_0 \} \cup \{ F_i | f_i \in F_{\!\cal P} \}$ is the stack alphabet,
$Z_0$ is the initial stack symbol, and
$\delta$ is the transition function defined as follows:
\vspace*{-4mm}

\begin{quote}
\noindent \normalsize
\begin{enumerate}
\item
$\delta(q_0,f_i,Z_0) = (q_F, F_i Z_0)$, \hspace*{3,05mm} $\forall f_i \in F_{\!\cal P}$,
\item
$\delta(q_F,f_i,F_j) = (q_F, F_i F_j)$, \hspace*{2,8mm} $\forall f_i \in F_{\!\cal P}$,
\item
$\delta(q_F,\overline{f}_j,F_j) = (q_F, \epsilon)$, \hspace*{7,8mm}\ $\forall f_i \in F_{\!\cal P}$.
\end{enumerate}
\end{quote}

\vspace*{1mm}
The input string is recognized if after having scanned the entire string the
automaton is in state $q_F$ and the stack contains at least one symbol $F_i$.

\begin{quote}
 \normalsize
A path $\pi$ is called:
\begin{itemize}
\item
\emph{increasing}, if $\hat{\lambda}(\pi) \in \LL(\Gamma_{\!\cal P})$,
\item
\emph{flat}, if $\hat{\lambda}(\pi) = \epsilon$,
\item
\emph{failing}, otherwise.
\end{itemize}
\end{quote}
\vspace{1mm}
It is worth noting that $\lambda(\pi) \in \LL(\Gamma_{\!\cal P})$ iff $\hat{\lambda}(\pi) \in \LL(\Gamma_{\!\cal P})$
as function $\hat{\lambda}$ emulates the pushdown automaton used to recognize
$\LL(\Gamma_{\!\cal P})$.
More specifically, for any path $\pi$ and relative string $\lambda(\pi)$ we have that:
\vspace*{-4mm}
\begin{quote}
\noindent \normalsize
\begin{itemize}
\item
if $\pi$ is increasing, then the pushdown automaton recognizes the string
$\lambda(\pi)$ in state $q_F$ and the stack contains a sequence of symbols corresponding to the symbols in $\hat{\lambda}(\pi)$ plus the initial stack symbol $Z_0$;
\item
if $\pi$ is flat, then the pushdown automaton does not recognize the string
$\lambda(\pi)$; moreover, 
the entire input string is scanned,
but the stack contains only the symbol $Z_0$;
\item
if $\hat{\lambda}(\pi)$ is failing, then the pushdown automaton does not recognize the string
$\lambda(\pi)$ as it goes in an error state.
\end{itemize}
\end{quote}

\paragraph{\bf Complexity.}
Concerning the complexity of checking whether a program is \acyclic,
we first introduce definitions and results
that will be used hereafter.
We start by introducing the notion of size of a logic program.

We assume that simple terms have
constant size and, therefore, the size of a complex term
$f(t_1,\dots,t_k)$, where $t_1,\dots,t_k$ are simple terms, is bounded
by $O(k)$. Analogously, the size of an atom $p(t_1,\dots,t_n)$ is given by the sum of
the sizes of the $t_i$'s, whereas the size of a conjunction of
atoms (resp. rule, program) is given by the sum of the sizes of its
atoms. That is, we identify for a program $\PP$ the following parameters:
$n_r$ is the number of rules of $\PP$,
$n_b$ is the maximum number of atoms in the body of rules of $\PP$,
$a_p$ is the maximum arity of predicate symbols occurring in $\PP$, and
$a_f$ is the maximum arity of function symbols occurring in $\PP$.
We assume that the size of $\PP$, denoted by $size(\PP)$, is bounded by $O(n_r \times n_b \times a_p \times a_f)$.
Finally, since checking whether a program is terminating requires to read the program,  we assume that the program has been already scanned and stored using suitable data structures.
Thus, all the complexity results presented in the rest of the paper do not take into account the cost of
scanning and storing the input program.
We first introduce a tighter bound for the complexity of computing $AR(\PP)$.

\begin{proposition}\label{prop:numberIter}
For any program $\PP$, the time complexity of computing $AR(\PP)$ is bounded by $O(|args(\PP)|^3)$.
\end{proposition}
\vspace*{-3mm}
\proof{
Assume that $n = |args(\PP)|$ is the total number of arguments of $\PP$.
First, it is important to observe the connection between the behavior
of operator $\Omega$ and the structure of the labeled argument graph $\G_L(\PP)$.
In particular, if the applications of the operator $\Omega$ change the rank of an argument $q[i]$  from $0$ to $k$,
then there is a path from an argument to $q[i]$ in $\G_L(\PP)$, where the number of edges labeled
with some positive function symbol minus the number of edges labeled
with some negative function symbol is at least $k$.
Given a cycle in a labeled argument graph, let us call it \emph{affected} if the number of edges labeled
with some positive function symbol is greater than the number of edges labeled
with some negative function symbol.

If an argument is not restricted, it is involved
in or depends on an affected cycle.
On the other hand, if after an application of $\Omega$ the rank assigned to an argument
exceeds $n$, this argument is not restricted \cite{LierlerL09}.
Recall that we are assuming that $d_{max} = 1$ and, therefore, $M = n \times d_{max} = n$.

Now let us show that after $2n^2+n$ iterations of $\Omega$  all not restricted arguments
exceed rank $n$.
Consider  an affected cycle and suppose that it contains $k$ arguments,
whereas the number of arguments depending on this cycle, but not belonging to it is $m$.
Obviously, $k+m\leq n$. 
All arguments involved in this cycle change their rank by at least
one after $k$ iterations of $\Omega$. Thus their ranks will be
greater than $n+m$ after $(n+m+1)*k$ iterations. The arguments
depending on this cycle, but not belonging to it, need at
most another $m$ iterations to reach the rank greater than $n$.
Thus all unrestricted arguments exceed the rank $n$ in
$(n+m+1)*k+m$ iterations of $\Omega$. Since $(n+m+1)*k+m=
nk+mk+(k+m) \leq 2n^2+n$, the restricted arguments are those
that at step $2n^2+n$ do not exceed rank $n$. It follows
that the complexity of computing $AR(\PP)$ is bounded by
$O(n^3)$ because we have to do $O(n^2)$ iterations and, for
each iteration we have to check the rank of $n$
arguments.~\hfill $\Box$
}
\\

In order to study the complexity  of computing \acyclic\ arguments
of a program we introduce a directed (not labeled) graph obtained from the propagation graph.

\begin{definition}[Reduction of $\Delta(\PP)$]\label{def:reduction}
Given a program $\PP$, the \emph{reduction} of $\Delta(\PP)$ is a  directed graph $\Deltac(\PP)$
whose nodes are the arguments of $\PP$ and
there is an edge $(p[i],q[j])$ in $\Deltac(\PP)$ iff
there is a path $\pi$ from $p[i]$ to $q[j]$ in $\Delta(\PP)$ such  that $\hat{\lambda}(\pi) \in F_{\!\cal P}$.~\hfill $\Box$
\end{definition}

\begin{figure}
  \includegraphics[width=7.5cm]{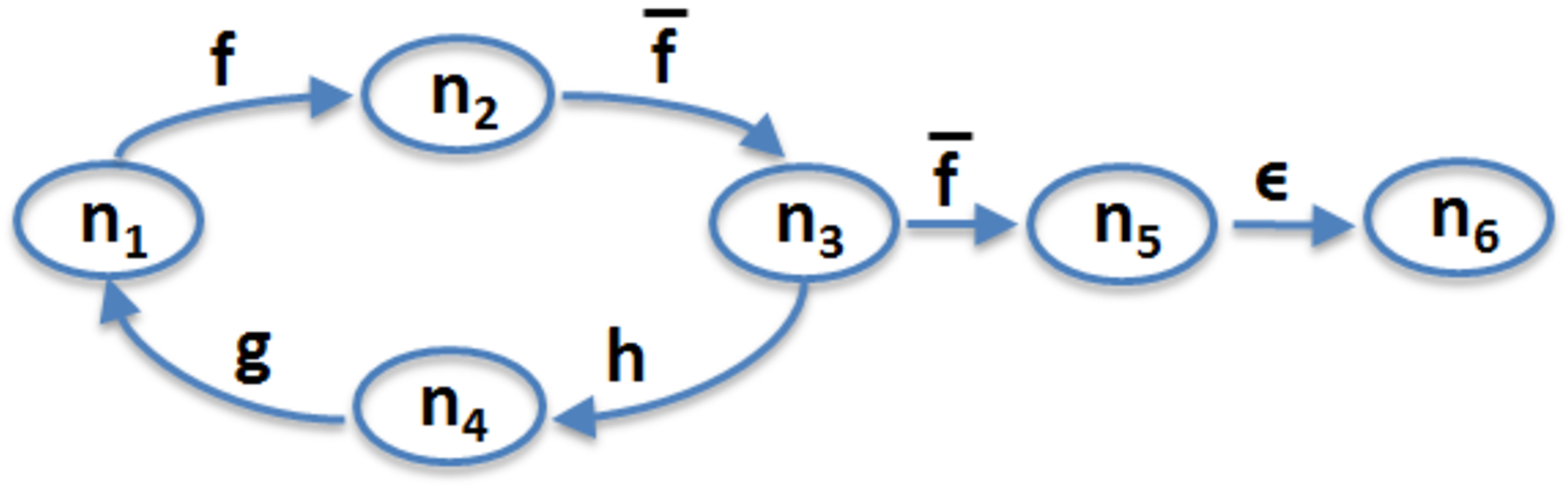}\\
  \caption{Propagation graph $\Delta(\PP)$ }\label{reduction1}
\end{figure}

\begin{figure}
  \includegraphics[width=7.5cm]{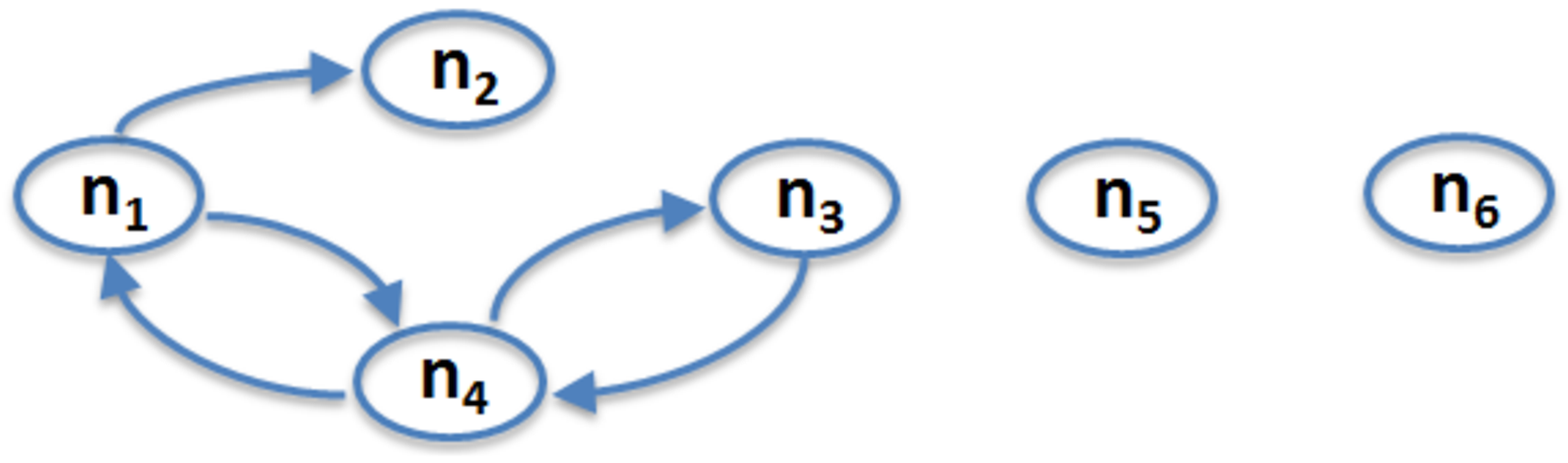}\\
  \caption{Reduction $\Deltac(\PP)$ of propagation graph $\Delta(\PP)$ }\label{reduction2}
\end{figure}

The reduction $\Deltac(\PP)$ of the propagation graph $\Delta(\PP)$ from Figure \ref{reduction1} is shown in Figure~\ref{reduction2}.
It is simple to note that for each path in $\Delta(\PP)$ from node $p[i]$ to node $q[j]$ spelling a string of $\LL(\Gamma_{\!\cal P})$
there exists a path from $p[i]$ to $q[j]$ in $\Deltac(\PP)$ and vice versa. As shown in the lemma below, this property always holds.

\begin{lemma}\label{lemma:cycles_equivalent}
Given a program $\PP$ and arguments $p[i],q[j]\in args(\PP)$, there exists a path in $\Delta(\PP)$ from $p[i]$ to $q[j]$ spelling a string of $\LL(\Gamma_{\!\cal P})$
iff there is a path from $p[i]$ to $q[j]$ in $\Deltac(\PP)$.
\end{lemma}
\vspace*{-3mm}
\proof{
($\Rightarrow$) 
Suppose there is a path $\pi$ from $p[i]$ to $q[j]$ in $\Delta(\PP)$ such that $\lambda(\pi) \in \LL(\Gamma_{\!\cal P})$.
Then $\hat{\lambda}(\pi)$ is a non-empty string, say $f_1 \dots f_k$, where $f_i\in F_{\cal P}$ for $i \in [1..k]$. Consequently, $\pi$ can be seen as a sequence of subpaths $\pi_1, \dots, \pi_k$, such that
$\hat{\lambda}(\pi_i)=f_i$ for $i \in [1..k]$.
Thus, from the definition of the reduction of $\Delta(\PP)$, there is a path from $p[i]$ to $q[j]$ in $\Deltac(\PP)$ whose length is equal to $|\hat{\lambda}(\pi)|$.

\noindent
($\Leftarrow$) Suppose there is a path $(n_1,n_2)\dots (n_k, n_{k+1})$ from $n_1$ to $n_{k+1}$ in $\Deltac(\PP)$. From the definition of the reduction of $\Delta(\PP)$,
for each edge $(n_i, n_{i+1})$ there is a path, say $\pi_i$, from $n_i$ to $n_{i+1}$ in $\Delta(\PP)$
such that $\hat{\lambda}(\pi_i) \in F_{\cal P}$.
Consequently, there is a path from $n_1$ to $n_{k+1}$ in $\Delta(\PP)$, obtained as a sequence of paths $\pi_1, \dots, \pi_k$ whose string is simply $\lambda(\pi_1) \dots \lambda(\pi_k)$.
Since $\hat{\lambda}(\pi_i) \in F_{\cal P}$ implies that $\lambda(\pi_i) \in \LL(\Gamma_{\!\cal P})$, for every $1 \le i \le k$, we have that
$\lambda(\pi_1) \dots \lambda(\pi_k)$ belongs also to $\LL(\Gamma_{\!\cal P})$.~\hfill $\Box$
}

\begin{proposition}\label{prop:complexity_reduction}
Given a program $\PP$, the time complexity of computing the reduction $\Deltac(\PP)$ is bounded by $O(|args(\PP)|^3 \times |F_{\!\PP}|)$.
\end{proposition}
\vspace*{-3mm}
\proof{
The construction of $\Deltac(\PP)$ can be performed as follows.
First, we compute all the paths $\pi$ in $\Delta(\PP)$ such that
$|\hat{\lambda}(\pi)| \leq 1$.
To do so, we use a slight variation of the Floyd-Warshall's transitive closure of $\Delta(\PP)$
which is defined by the following recursive formula.
Assume that each node of $\Delta(\PP)$ is numbered from $1$ to $n=|args(\PP)|$, then
we denote with $path(i,j,\alpha,k)$ the existence of a path $\pi$ from node $i$ to node $j$ in $\Delta(\PP)$
such that $\hat{\lambda}(\pi) = \alpha$, $|\alpha| \leq 1$
and $\pi$ may go only through nodes in $\{1,\dots,k\}$ (except for $i$ and $j$).

The set of atoms $path(i,j,\alpha,k)$, for all values $1 \leq i,j \leq n$, can be derived
iteratively as follows:
\vspace*{-5mm}
\begin{quote}
\noindent \normalsize
\begin{itemize}
\item (base case: $k=0$)
$path(i,j,\alpha,0)$ holds if there is an edge $(i,j,\alpha)$ in $\Delta(\PP)$,
\item (inductive case: $0<k\leq n$)
$path(i,j,\alpha,k)$ holds if
\begin{itemize}
\item
$path(i,j,\alpha,k-1)$ holds, or
\item
$path(i,k,\alpha_1 ,k-1)$ and  $path(k,j,\alpha_2 ,k-1)$ hold, $\alpha = \alpha_1 \alpha_2$ and $|\alpha| \leq 1$.
\end{itemize}
\end{itemize}
\end{quote}

Note that in order to compute all the possible atoms $path(i,j,\alpha,k)$, we need to first
initialize every base atom $path(i,j,\alpha,0)$ with cost bounded by $O(n^2 \times |F_{\!\PP}|)$,
as this is the upper bound for the number of edges in $\Delta(\PP)$.
Then, for every $1 \le k \le n$, we need to compute all paths $path(i,j,\alpha,k)$, thus requiring
a cost bounded by $O(n^3 \times |F_{\!\PP}|)$ operations.
The whole procedure will require $O(n^3 \times |F_{\!\PP}|)$ operations.
Since we have computed all possible paths $\pi$ in $\Delta(\PP)$ such that $|\hat{\lambda}(\pi)| \leq 1$,
we can obtain all the edges $(i,j)$ of $\Deltac(\PP)$ (according to Definition~\ref{def:reduction}) by simply
selecting the atoms $path(i,j,\alpha,k)$ with $\alpha \in F_{\!\PP}$, whose cost is bounded
by $O(n^2 \times |F_{\!\PP}|)$.
Then, the time complexity of constructing $\Deltac(\PP)$ is $O(n^3 \times |F_{\!\PP}|)$.~\hfill $\Box$
}

\begin{theorem}\label{acyclic-decidable}
The complexity of deciding whether a program $\PP$ is
\acyclic\ is bounded by $O(|args(\PP)|^3 \times |F_{\!\cal P}|)$.
\end{theorem}
\vspace*{-3mm}
\proof{
Assume that $n = |args(\PP)|$ is the total number of arguments of $\PP$. To check whether $\PP$ is \acyclic\ it is sufficient to
first compute the set of restricted arguments $AR(\PP)$ which requires time $O(n^3)$ from Proposition~\ref{prop:numberIter}.
Then, we need to construct the propagation graph $\Delta(\PP)$, for which the maximum number of edges is $n^2 \times ( |F_{\!\PP}| + |\overline{F}_{\!\PP}| + 1)$,
then it can be constructed in time $O(n^2 \times |F_{\!\PP}|)$ (recall that we are not taking into account
the cost of scanning and storing the program). Moreover, starting from $\Delta(\PP)$, we need to construct $\Deltac(\PP)$, which requires time $O(n^3 \times |F_{\!\PP}|)$
(cf. Proposition~\ref{prop:complexity_reduction}) and
then, following Lemma~\ref{lemma:cycles_equivalent}, we need to check whether $\Deltac(\PP)$ is acylic. Verifying whether $\Deltac(\PP)$ is acyclic can be done by means of a simple traversal
of $\Deltac(\PP)$ and checking if a node is visited more than once.
The complexity of a depth-first traversal of a graph is well-known to be $O(|E|)$ where $E$ is the set of edges of the graph.
Since the maximum number of edges of $\Deltac(\PP)$ is by definition $n^2 \times |F_{\!\PP}|$,
the traversal of $\Deltac(\PP)$ can be done in time $O(n^2 \times |F_{\!\PP}|)$.
Thus, the whole time complexity is still bounded by $O(n^3 \times |F_{\!\PP}|)$.~\hfill $\Box$
}

\vspace*{3mm}
\begin{corollary}\label{coro:numberIter}
For any program $\PP$, the time complexity of computing $\GA(\PP)$ is bounded by $O(|args(\PP)|^3 \times |F_{\!\PP}|)$.
\end{corollary}
\vspace*{-3mm}
\proof{
Straightforward from the proof of Theorem \ref{acyclic-decidable}.
\hfill $\Box$
}

\vspace*{3mm}

As  shown in the previous theorem, the time complexity of checking whether a program $\PP$
is \acyclic\ is bounded by $O(|args(\PP)|^3 \times |F_{\!\PP}|)$,
which is strictly related to the complexity of checking whether a program is argument-restricted,
which is $O(|args(\PP)|^3)$. In fact,
the new proposed criterion performs a more accurate analysis on how terms are propagated from the body to the head of rules by taking into account
the function symbols occurring in such terms. Moreover,
if a logic program $\PP$ has only one function symbol, the time complexity
of checking whether $\PP$ is \acyclic\ is the same as the one required to check if it is argument-restricted.

\section{Safe programs}\label{sec:safe}

The \acyclicity\ termination criterion presents some limitations, since it is not able to detect when
a rule can be activated only a finite number of times during the bottom up evaluation of the program.
The next example shows a simple terminating program which is not recognized by the \acyclicity\ termination criterion.

\begin{example}\label{Example-act-graph}
Consider the following logic program $P_{\ref{Example-act-graph}}$:
\[
\begin{array}{l}
r_1: \tt p(X,X) \leftarrow b(X). \\
r_2: \tt p(f(X),g(X)) \leftarrow p(X,X)\. \\
\end{array}
\]
where $\tt b$ is base predicate.
As the program is standard, it has a (finite) unique minimal model, which can can be derived
using the classical bottom-up fixpoint computation algorithm.
Moreover, independently from the set of base facts defining $\tt b$,
the minimum model of $P_{\ref{Example-act-graph}}$ is finite and its computation terminates.
~\hfill $\Box$
\end{example}

Observe that the rules of program $P_{\ref{Example-act-graph}}$ can be activated at most $n$ times,
where $n$ is the cardinality of the active domain of the base predicate $\tt b$.
Indeed, the recursive rule $r_2$ cannot activate itself since the
newly generated atom is of the form $\tt p(f(\cdot),g(\cdot))$ and does not unify with its body.\\
As another example consider the recursive rule $\tt q(f(X)) \leftarrow q(X), t(X)$
and the strongly linear rule $\tt p(f(X), g(Y)) \leftarrow p(X,Y), t(X)$ where $\tt t[1]$ is a limited  argument. 
The activation of these rules is limited by the cardinality of the active domain of $\tt t[1]$.

\nop{As another example consider a rule $\tt p(f(X)) \leftarrow p(X), b(X)$ where $\tt b$ is base predicate. This rule can be executed only a finite number of times since $\tt b[1]$ is limited. Finally, observe that also the strongly recursive rule
$\tt p(f(X), g(Y)) \leftarrow p(X,Y), b(X)$ where $\tt b$ is base predicate
can be executed only a finite number of times.}

Thus, in this section, in order to define a more general termination criterion
we introduce the \emph{safety} function which, by detecting rules that can be executed only a finite number of times,
derives a larger set of limited arguments of the program.
We start by analyzing how rules may activate each other.

\begin{definition}[Activation Graph]\label{actvGraph}
Let $\PP$ be a program and let $r_1$ and $r_2$ be (not necessarily distinct) rules of $\PP$.
We say that $r_1$ \emph{activates} $r_2$  iff
$head(r_1)$ and an atom in $body(r_2)$ unify.
The \emph{activation graph} $\Sigma(\PP)=(\PP,E)$ consists of the set of nodes denoting the rules
of $\PP$ and the set of  edges $(r_i,r_j)$, with $r_i,r_j \in \PP$, such that $r_i$ activates $r_j$.
\hfill $\Box$
\end{definition}

\begin{example}
Consider program $P_{\ref{Example-act-graph}}$ of Example \ref{Example-act-graph}.
The  activation graph of this program contains two nodes $r_1$ and $r_2$ and an edge from $r_1$ to $r_2$.
Rule $r_1$ activates rule $r_2$ as
the head atom $\tt p(X,X)$ of $r_1$ unifies with the body atom $\tt p(X,X)$ of $r_2$.
Intutively, this means that the execution of the first rule may cause the second rule to be activated. In fact, the execution of
$r_1$ starting from the database instance $D=\tt \{b(a)\}$ produces the new atom $\tt p(a,a)$.
The presence of this atom allows the second rule to be activated, since the body of $r_2$ can be made true by means of the
atom $\tt p(a,a)$, producing the new atom $\tt p(f(a),g(a))$.
It is worth noting that the second rule cannot activate itself since $head(r_2)$ does not unify with the atom $\tt p(X,X)$ in $body(r_2)$.~\hfill $\Box$
\end{example}

The activation graph shows how rules may activate each other, and, consequently, the possibility to propagate values from one rule to another.
Clearly, the active domain of an argument $p[i]$ can be infinite only if $p$ is the head predicate of a rule that may be activated an infinite number of times.
A rule may be activated an infinite number
of times only if it depends on a cycle of the activation graph.
Therefore, a rule not depending on a cycle
can only propagate a finite number of values into its head arguments.
Another important aspect is the structure of rules and the presence of limited arguments in their body and head atoms.
As  discussed at the beginning of this section, rules $\tt q(f(X)) \leftarrow q(X), t(X)$ and  $\tt p(f(X), g(Y)) \leftarrow p(X,Y), t(X)$, where $\tt t[1]$ is a limited  argument, can be activated only a finite number of times.
In fact, as variable $\tt X$ in both rules can be substituted only by values taken from the active domain of $\tt t[1]$, which is finite, the active domains of $\tt q[1]$ and $\tt p[1]$ are finite as well, i.e. $\tt q[1]$ and $\tt p[1]$ are limited arguments. Since $\tt q[1]$ is limited, the first rule can be applied only a finite number of times. In the second rule we have predicate $\tt p$ of arity two in the head, and we know that $\tt p[1]$ is a limited argument.
Since the second rule is strongly linear, the domains of both head arguments
$\tt p[1]$ and $\tt p[2]$ grow together each time this rule is applied.
Consequently, the active domain of $\tt p[2]$ must be finite as well as the active domain of $\tt p[1]$
and this rule can be applied only a finite number of times.

\vspace*{3mm}
We now introduce the notion of \emph{limited term}, that will be used to define a function, called \emph{safety function}, that takes as input a set of limited arguments  and derives a new set of limited arguments in $\PP$.

\begin{definition}[Limited terms]\label{limited-term-def}
Given a rule $r = q(t_1,\dots,t_m) \leftarrow body(r) \in \PP$ and a set $A$ of limited arguments,
we say that $t_i$ is \emph{limited} in $r$ (or $r$ \emph{limits} $t_i$) w.r.t. $A$
if one of the following conditions holds:
\begin{enumerate}
 \item every variable $X$ appearing in $t_i$ also appears in an argument in
$body(r)$ belonging to $A$, or
\item
$r$ is a strongly linear rule such that:
\begin{enumerate}
\item
for every atom $p(u_1,...,u_n) \in head(r) \cup rbody(r)$,\\ all terms $u_1,...,u_n$ are either simple or complex;
\item
$var(head(r)) = var(rbody(r))$,
\item
there is an argument $q[j] \in A$.
\hfill $\Box$
\end{enumerate}
\end{enumerate}
\end{definition}

\begin{definition}[Safety Function]\label{safe-function}
For any program $\PP\!$, let $A$ be a set of limited arguments of $\PP\!$ and let $\Sigma(\PP)$
be the activation graph of $\PP$.
The \emph{safety function} $\Psi(A)$ denotes the set of arguments $q[i] \in args(\PP)$
such that for all rules $r = q(t_1,\dots,t_m) \leftarrow body(r) \in \PP$, either
$r$ does not depend on a cycle $\pi$ of $\Sigma(\PP)$ or $t_i$ is limited in $r$ w.r.t. $A$.
\hfill~$\Box$
\end{definition}

\begin{figure}
  \includegraphics[width=9cm]{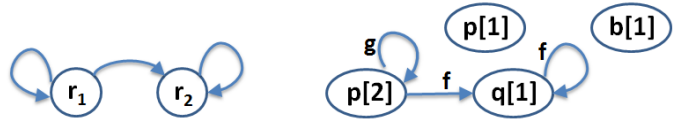}\\
  \caption{Activation (left) and propagation (right) graphs of program $P_{\ref{Example-safety-function}}$. }\label{FigureEs12}
\end{figure}

\begin{example}\label{Example-safety-function}
 Consider the following program $P_{\ref{Example-safety-function}}$:
\[
\begin{array}{l}
r_1\!:\ \tt p(f(X),g(Y)) \leftarrow p(X,Y), b(X)\. \\
r_2\!:\ \tt q(f(Y)) \leftarrow p(X,Y), q(Y)\.
\end{array}
\]
where $\tt b$ is base predicate. Let $A=\GA(\PP)=\{\tt b[1],p[1]\}$. The activation and the propagation graphs
of this program are reported in Figure~\ref{FigureEs12}.
The application of the safety function to the set of limited arguments $A$ gives $\Psi(A)=\{\tt b[1],p[1], p[2]\}$.
Indeed:
\begin{itemize}
\item
$\texttt{b[1]} \in \Psi(A)$ since $\tt b$ is a base predicate which does not appear in the head of any rule;
consequently all the rules with $\tt b$ in the head (i.e. the empty set)
trivially satisfy the conditions of Definition~\ref{safe-function}.
\item
$\texttt{p[1]} \in \Psi(A)$ because the unique rule with $\tt p$ in the head (i.e. $r_1$)
satisfies the first condition of Definition~\ref{limited-term-def}, that is, $r_1$ limits the term $\tt f(X)$ w.r.t. $A$
in the head of rule $r_1$ corresponding to argument $\tt p[1]$.
\item
Since $r_1$ is strongly linear and the second condition of Definition~\ref{limited-term-def} is satisfied,
$\texttt{p[2]} \in \Psi(A)$ as well.
\hfill~$\Box$
\end{itemize}
\end{example}

\vspace*{3mm}

The following proposition shows that the safety function can be used to derive further
limited arguments.

\begin{proposition}\label{prop:safe-func-fd}
Let $\PP$ be a program and let $A$ be a set of  limited arguments
of $\PP$. Then, all arguments in $\Psi(A)$ are also limited.
\end{proposition}
\vspace*{-3mm}
\proof{
Consider an argument $q[i] \in \Psi(A)$, then for every rule $r=q(t_1,\dots,t_n) \leftarrow body(r)$
either $r$ does not depend on a cycle of $\Sigma(\PP)$ or $t_i$ is limited in $r$ w.r.t. $A$.

Clearly, if $r$ does not depend on a cycle of $\Sigma(\PP)$, it can be activated a finite number of times
as it is not 'effectively recursive' and does not depend on rules which are effectively recursive.

Moreover, if $t_i$ is limited in $r$ w.r.t. $A$, we have that either:\\
\emph{1)}
The first condition of Definition \ref{limited-term-def} is satisfied (i.e. every
variable $X$ appearing in $t_i$ also appears in an argument in
$body(r)$ belonging to $A$).
This means that variables in $t_i$ can be replaced by a finite number of values. \\
\emph{2)}
The second condition of Definition \ref{limited-term-def} is satisfied.
Let $p(t_1,...,t_n) = head(r)$,
the condition that all terms $t_1,...,t_n$ must be simple or complex guarantees
that, if terms in $head(r)$ grow, then they grow all together (conditions 2.a and 2.b).
Moreover, if the growth of a term $t_j$ is blocked (Condition 2.c),
the growth of all terms (including $t_i$) is blocked too.    \\
Therefore, if one of the two condition is satisfied for all rules defining $q$,
the active domain of $q[i]$ is finite.
~\hfill $\Box$
}
\\

Unfortunately, as shown in the following example, the relationship $A \subseteq \Psi(A)$
does not always hold for a generic set of arguments $A$, even if the arguments in $A$ are limited.

\begin{figure}
  \includegraphics[width=6cm]{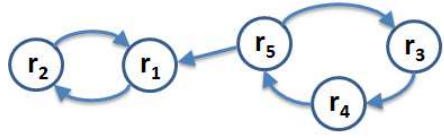}\\
  \caption{Activation Graph of program $P_{\ref{Example-notsafe-AR}}$ }\label{FigureActGraphsEs11}
\end{figure}

\begin{example}\label{Example-notsafe-AR}
Consider the following program $P_{\ref{Example-notsafe-AR}}$:
\[
\begin{array}{l}
r_1: \tt p(f(X),Y) \leftarrow q(X),r(Y)\. \\
r_2: \tt q(X) \leftarrow p(X,Y)\.   \\
r_3: \tt t(Y) \leftarrow r(Y)\.\\
r_4: \tt s(Y) \leftarrow t(Y)\.\\
r_5: \tt r(Y) \leftarrow s(Y)\.
\end{array}
\]
Its activation graph $\Sigma(P_{\ref{Example-notsafe-AR}})$ is shown in Figure~\ref{FigureActGraphsEs11}, whereas the set of restricted arguments is $AR(P_{\ref{Example-notsafe-AR}}) =
\GA(P_{\ref{Example-notsafe-AR}}) = \tt \{ r[1], t[1],$ $\tt s[1], p[2] \}$.
Considering the set $A = \tt \{ p[2] \}$, we have that the safety
function $\Psi({\tt \{p[2]\}}) = \emptyset$.
Therefore, the relation $A \subseteq \Psi(A)$ does not hold for $A =\tt \{p[2]\}$.

Moreover, regarding the set $A'= \GA(P_{\ref{Example-notsafe-AR}}) =\tt \{r[1],t[1],s[1],p[2]\}$, we have
$\Psi(A') =\tt \{r[1],t[1],$ $\tt s[1],p[2]\}$ $=A'$,
i.e. the  relation $A' \subseteq \Psi(A')$ holds.~\hfill $\Box$
\end{example}

\vspace*{2mm}
The following proposition states that  if we consider the set $A$ of \acyclic\ arguments
of a given program $\PP$, the relation $A \subseteq \Psi(A)$ holds.

\vspace*{2mm}
\begin{proposition}\label{FD:prop}
For any logic program $\PP$:
\begin{enumerate}
\item
$\GA(\PP) \subseteq \Psi(\GA(\PP))$;
\item
$\Psi^i(\GA(\PP)) \subseteq \Psi^{i+1}(\GA(\PP))$ for $i >0$.
\end{enumerate}
\end{proposition}
\vspace*{-3mm}
\proof{
\emph{1)}
Suppose that $q[k] \in \GA(\PP)$.
Then  $q[k]\in AR(\PP)$ or $q[k]$ does not
depend on a cycle in $\Delta(\PP)$ spelling a string of
$\LL(\Gamma_{\!\cal P})$. In both cases $q[k]$ can depend only on
arguments in $\GA(\PP)$. If  $q[k]$ does not depend on any argument,
then it does not appear in the head of any rule and, consequently,
$q[k] \in \Psi(\GA(\PP))$. Otherwise, the first condition of
Definition~\ref{limited-term-def} is satisfied and $q[k] \in
\Psi(\GA(\PP))$.
\\
\emph{2)}
We prove that $\Psi^i(\GA(\PP)) \subseteq \Psi^{i+1}(\GA(\PP))$ for $i >0$
by induction.
We start by showing that $\Psi^i(\GA(\PP)) \subseteq \Psi^{i+1}(\GA(\PP))$ for $i=1$, i.e. that the relation $\Psi(\GA(\PP)) \subseteq \Psi(\Psi(\GA(\PP)))$ holds.
In order to show this relation
we must show that for every argument $q[k]\in \PP$ if $q[k] \in \Psi(\GA(\PP))$,
then $q[k] \in \Psi(\Psi(\GA(\PP))$.
Consider $q[k] \in \Psi(\GA(\PP))$. Then, $q[k]$ satisfies  Definition~\ref{safe-function} w.r.t. $A=\GA(\PP)$.
From comma one of this proof it follows that
$\GA(\PP) \subseteq \Psi(\GA(\PP))$, consequently
$q[k]$ satisfies Definition~\ref{safe-function} w.r.t. $A=\Psi(\GA(\PP))$ too and so, $q[k] \in \Psi(\Psi(\GA(\PP)))$.

Suppose that $\Psi^k(\GA(\PP)) \subseteq \Psi^{k+1}(\GA(\PP))$ for $k >0$.
In order to show that $\Psi^{k+1}(\GA(\PP)) \subseteq \Psi^{k+2}(\GA(\PP))$
we must show that for every argument $q[k]\in \PP$ if $q[k] \in \Psi^{k+1}(\GA(\PP))$,
then $q[k] \in \Psi^{k+2}(\GA(\PP))$.
Consider $q[k] \in \Psi^{k+1}(\GA(\PP))$. Then $q[k]$ satisfies Definition~\ref{safe-function} w.r.t. $A=\Psi^k(\GA(\PP))$.
Since $\Psi^k(\GA(\PP)) \subseteq \Psi^{k+1}(\GA(\PP))$,
$q[k]$ satisfies Definition~\ref{safe-function} w.r.t. $A=\Psi^{k+1}(\GA(\PP))$ too. Consequently, $q[k] \in \Psi^{k+2}(\GA(\PP))$.
~\hfill$\Box$

}

\vspace*{2mm}
Observe that we can prove in a similar way that
$AR(\PP) \subseteq \Psi(AR(\PP))$ and
that $\Psi^i(AR(\PP)) \subseteq \Psi^{i+1}(AR(\PP))$ for $i >0$.

\vspace*{2mm}
\begin{definition}[Safe Arguments and Safe Programs]\label{affected-arguments}
For any program $\PP$, $\safe(\PP) = \Psi^\infty(\GA(\PP))$ denotes the set of \emph{safe arguments} of $\PP$.
A program $\PP$ is said to be \emph{safe} if all arguments are safe.
The class of safe programs will be denoted by $\Safe$.
\hfill $\Box$
\end{definition}

\vspace*{2mm}

Clearly, for any set of arguments $A \subseteq \GA(\PP)$,
$\Psi^i(A) \subseteq \Psi^i(\GA(\PP))$.
Moreover, as shown in Proposition~\ref{FD:prop}, when the starting set is $\GA(\PP)$, the sequence $\GA(\PP),\ \Psi(\GA(\PP)),\Psi^2(\GA(\PP)),\ \dots$ is monotone and
there is a finite $n = O(|args(\PP)|)$ such that $\Psi^n(\GA(\PP))  = \Psi^\infty(\GA(\PP))$.
We can also define the inflactionary version of $\Psi$ as
$\hat{\Psi}(A) = A \cup \Psi(A)$, obtaining that
$\hat{\Psi}^i(\GA(\PP))  = \Psi^i(\GA(\PP))$, for all natural numbers $i$.
The introduction of the inflactionary version
guarantees that the sequence $A,\ \hat{\Psi}(A),\ \hat{\Psi}^2(A),\ \dots$ is monotone
for every
set $A$ of limited arguments.
This would allow us to derive a (possibly) larger set of limited arguments starting from
any set of limited arguments.

\begin{example}\label{Example-safe}
Consider again program $P_{\ref{Example-act-graph}}$ of Example
\ref{Example-act-graph}.
Although $AR(P_{\ref{Example-act-graph}})=\emptyset$, the program $P_{\ref{Example-act-graph}}$
is safe as $\Sigma(P_{\ref{Example-act-graph}})$ is acyclic. \\
Consider now the program $P_{\ref{Example-safety-function}}$ of Example \ref{Example-safety-function}.
As already shown in Example \ref{Example-safety-function}, the first application of the safety function to the set of
$\Gamma$-acyclic arguments of $P_{\ref{Example-safety-function}}$ gives $\Psi(\GA(P_{\ref{Example-safety-function}}))=\{\tt b[1],p[1], p[2]\}$.
The application of the safety function to the obtained set gives
$\Psi(\Psi(\GA(P_{\ref{Example-safety-function}})))=\{\tt b[1],p[1], p[2],q[1]\}$.
In fact, in the unique rule defining $\tt q$, term $\tt f(Y)$, corresponding to the argument $\tt q[1]$,
is limited in $r$ w.r.t. $\{\tt b[1], p[1], p[2]\}$ (i.e. the variable $\tt Y$ appears in $body(r)$ in a term corresponding to argument $\tt p[2]$
and argument $\tt p[2]$, belonging to the input set, is limited).
At this point, all arguments of $P_{\ref{Example-safety-function}}$ belong to the resulting set. Thus, $\safe(P_{\ref{Example-safety-function}}) = args(P_{\ref{Example-safety-function}})$, and
we have that program $P_{\ref{Example-safety-function}}$ is
safe.~\hfill~$\Box$
\end{example}

We now show results on the expressivity of the class $\Safe$ of safe programs.

\begin{theorem}\label{theorem-safeFG}
The class $\Safe$ of safe programs strictly includes the class $\AC$ of \acyclic\ programs and is strictly contained in the class $\FG$ of finitely ground programs. 
\end{theorem}
\vspace*{-3mm}
\proof{
($\AC \subsetneq \Safe$).
From Proposition~\ref{FD:prop} it follows that $\AC \subseteq \Safe$.
Moreover, $\AC \subsetneq \Safe$ as program $P_{\ref{Example-safety-function}}$ is safe but not \acyclic.

\vspace*{3mm}
\noindent
($\Safe \subsetneq \FG$).
From Proposition~\ref{prop:safe-func-fd} it follows that
every argument in the safe program  can take values only from
a finite domain.
Consequently, the set of all possible ground terms derived
during the grounding process is finite and the program is finitely ground.
Moreover, we have that the program $P_{\ref{Example1-Intro}}$ of Example~\ref{Example1-Intro} is finitely ground, but not safe.
~\hfill$\Box$
}
\\

As a consequence of Theorem \ref{theorem-safeFG}, every safe program admits a finite minimum model.

\paragraph{\bf Complexity.}
We start by introducing a bound on the complexity of constructing the activation graph.

\begin{proposition}\label{activation-compl}
For any program $\PP$, the activation graph $\Sigma(\PP)$ can be
constructed in time 
$O(n_r^2 \times n_b \times (a_p \times a_f)^2)$,
where $n_r$ is the number of rules of $\PP$,
$n_b$ is the maximum number of body atoms in a rule,
$a_p$ is the maximum arity of predicate symbols and $a_f$ is the maximum arity of function symbols.
\end{proposition}
\vspace*{-3mm}
\proof{
To check whether a rule $r_i$ activates a rule $r_j$ we have to determine if an atom $B$ in $body(r_j)$ unifies with the head-atom $A$ of $r_i$.
This can be done in time $O(n_b \times u)$, where
$u$ is the cost of deciding whether two atoms unify, which is quadratic in the size of the two atoms \cite{Venturini75},
that is $u = O((a_p \times a_f)^2)$ as the size of atoms is bounded by $a_p \times a_f$
(recall that the maximum depth of terms is 1).
In order to construct the activation graph we have to consider all pairs of rules and for each pair we have to check if the first rule
activates the second one. Therefore, the global complexity is $O(n_r^2 \times n_b \times u)=O(n_r^2 \times n_b \times (a_p \times a_f)^2)$.
\hfill$\Box$
}

\vspace*{3mm}
We recall that given two atoms $A$ and $B$, the size of a m.g.u. $\theta$ for $\{A, B \}$ can be, in the worst case,
exponential in the size of $A$ and $B$, but the complexity of deciding whether a unifier for $A$ and $B$ exists
is quadratic in the size of $A$ and $B$ \cite{Venturini75}.

\begin{proposition}\label{theorem-recognizeSafe}
The complexity of deciding whether a program $\PP$ is safe is
$O((size(\PP))^2 + |args(\PP)|^3 \times |F_{\!\cal P}|)$. 
\end{proposition}
\vspace*{-3mm}
\proof{
The construction of the activation graph $\Sigma(\PP)$ can be done in
time  $O(n_r^2 \times n_b \times (a_p \times a_f)^2)$, where $n_r$ is the number of rules of $\PP$,
$n_b$ is the maximum number of body atoms in a rule,
$a_p$ is the maximum arity of predicate symbols and $a_f$ is the maximum arity of function symbols
(cf. Proposition \ref{activation-compl}).

The complexity of computing $\GA(\PP)$ is bounded by $O(|args(\PP)|^3 \times |F_{\!\cal P}|)$
(cf. Theorem \ref{acyclic-decidable}).

From Definition~\ref{safe-function} and
Proposition~\ref{FD:prop} it follows that the sequence $\GA(\PP)$,
$\Psi(\GA(\PP))$, $\Psi^2(\GA(\PP))$, \.\.\. is monotone and converges in a finite number of steps bounded
by the cardinality of the set $args(\PP)$.
The complexity of determining rules not depending on cycles in the activation graph $\Sigma(\PP)$
is bounded by $O(n_r^2)$, as it can be done by means of a depth-first traversal of $\Sigma(\PP)$,
which is linear in the number of its edges.
Since checking whether the conditions of Definition~\ref{limited-term-def} hold for all arguments
in $\PP$ is in $O(size(\PP))$, checking such conditions for at most $|args(\PP)|$ steps
is $O(|args(\PP)| \times size(\PP))$. Thus, the complexity of checking all the conditions of Definition~\ref{safe-function}
for all steps is $O(n_r^2 + |args(\PP)| \times size(\PP))$.

Since, $n_r^2 \times n_b \times (a_p \times a_f)^2 = O((size(\PP))^2)$,
$|args(\PP)| = O(size(\PP))$ and $n_r^2 = O((size(\PP))^2)$,
the complexity of deciding whether $\PP$ is safe is $O((size(\PP))^2+|args(\PP)|^3 \times |F_{\!\cal P}|)$.
~\hfill$\Box$
}


\section{Bound Queries and Examples}\label{sec:boundQuery}

In this section we consider the extension of our framework to queries.
This is an important aspect as in many cases, the answer to a query is
finite, although the models may have infinite cardinality. This happens very often when the query goal contains ground
terms.

\subsection{Bound Queries}

Rewriting techniques, such as magic-set,
allow bottom-up evaluators to efficiently compute (partially) ground queries,
that is queries whose query goal contains ground terms.
These techniques rewrite queries (consisting of a query goal and a program) such that the top-down evaluation is emulated \cite{Beeri-91,Greco03,GrecoGTZ05,AlvianoFL10}.
Labelling techniques similar to magic-set have been also studied in the context of term rewriting \cite{Zantema95}.
Before presenting the rewriting technique, let us introduce some notations.

A \emph{query} is a pair $Q = \<q(u_1,\.\.,u_n),\PP\>$, where $q(u_1,\.\.,u_n)$ is an atom called
\emph{query goal} and $\PP$ is a program.
We recall that an \emph{adornment} of a predicate symbol $p$ with arity $n$ is a string $\alpha \in \{b,f\}^*$
such that $|\alpha| = n$\footnote{Adornments of predicates, introduced to optimize the bottom-up
computation of logic queries, are similar to \emph{mode of usage} defined
in logic programming to describe how the arguments of a predicate $p$ must be restricted when an atom
with predicate symbol $p$ is called.}.
The symbols $b$ and $f$ denote, respectively, \emph{bound} and \emph{free} arguments.
Given a query  $Q = \<q(u_1,\.\.,u_n),\PP\>$,
$MagicS(Q) = \<q^\alpha(u_1,\.\.,u_n), MagicS(q(u_1,\.\.,u_n),\PP)\>$ indicates the rewriting
of $Q$, where $MagicS(q(u_1,\.\.,u_n),\PP)$ denotes the rewriting of rules in $\PP$
with respect to the query goal $q(u_1,\.\.,u_n)$ and $\alpha$ is the adornment
associated with the query goal.

We assume that our queries $\<G,\PP\>$ are positive, as the rewriting technique is
here applied to $\<G, st(\PP)\>$ to generate the positive program which is used to restrict the source
program (see Section \ref{sec:compSM}).

\begin{definition}\label{weakly-acyclic-program}
A query $Q = \<G,\PP\>$ is \emph{safe} if
$\PP$ or $MagicS(G,\PP)$ is safe.~\hfill $\Box$
\end{definition}

It is worth noting that it is possible to have a
query $Q$=$\<G,\PP\>$ such that $\PP$ is safe, but the rewritten
program $MagicS(G,\PP)$ is not safe and vice versa.

\begin{example}\label{example-query}
Consider the query $ Q = {\tt \<p(f(f(a)))},P_{\ref{example-query}}\>$,
where $P_{\ref{example-query}}$ is defined  below:
\[
\begin{array}{l}
\tt p(a)\. \\
\tt p(f(X)) \!\leftarrow\! p(X)\.
\end{array}
\]
$P_{\ref{example-query}}$ is not safe, but if we rewrite the program
using the magic-set method, we obtain the safe program:
\[
\begin{array}{lll}
\tt magic\_p^{b}(f(f(a)))\. \\
\tt magic\_p^b(X) \leftarrow magic\_p^{b}(f(X))\. \smallskip \\
\tt p^{b}(a)     \leftarrow magic\_p^{b}(a)\. \\
\tt p^{b}(f(X))  \leftarrow magic\_p^{b}(f(X)),\ p^{b}(X)\. \\
\end{array}
\]

\vspace*{2mm}
\noindent
Consider now the query $Q = {\tt
\<p(a)},\PP'_{\ref{example-query}}\>$, where $\PP'_{\ref{example-query}}$ is defined as follows:
\[
\begin{array}{l}
\tt p(f(f(a)))\. \\
\tt p(X) \!\leftarrow\! p(f(X))\.
\end{array}
\]
The program is safe, but after the magic-set rewriting
we obtain the following program:
\[
\begin{array}{l}
\tt magic\_p^{b}(a)\. \\
\tt magic\_p^{b}(f(X)) \leftarrow magic\_p^{b}(X)\. \\
\tt p^{b}(f(f(a)))     \leftarrow magic\_p^{b}(f(f(a)))\. \\
\tt p^{b}(X)  \leftarrow magic\_p^{b}(X), p^{b}(f(X))\.
\end{array}
\]
which is not recognized as safe because it is not terminating.~\hfill $\Box$
\end{example}

Thus, we propose to first check if the input program is
safe and, if it does not satisfy the safety criterion,
to check the property on the rewritten program, which is query-equivalent
to the original one. \\
We recall that for each predicate symbol $p$ with arity $n$, the number of adorned predicates $p^{\alpha_1...\alpha_n}$ could be exponential and bounded by $O(2^n)$. However, in practical cases only few adornments are generated for each predicate symbol. 
Indeed, rewriting techniques are well consolidated and widely used to compute bound queries.

\subsection{Examples}

Let us now consider the application of the technique described above to some practical examples.
Since each predicate in the rewritten query has a unique adornment, we shall omit them.

\begin{example}\label{reverse-example}
Consider the query $\< {\tt reverse([a,b,c,d],L)}, P_{\ref{reverse-example}}\>$, where $P_{\ref{reverse-example}}$
is defined by the following rules:
\[
\begin{array}{l}
r_0:\tt\ reverse([\,], [\,])\. \\
r_1:\tt\ reverse([X|Y],[X|Z]) \leftarrow reverse(Y,Z)\. \ \ \ \ \ \ \ \ \ \ \ \ \ \ \ \ \\
\end{array}
\]
The equivalent program $P'_{\ref{reverse-example}}$, rewritten to be computed by means of a bottom-up evaluator, is:
\[
\begin{array}{l}
\rho_0:\tt\ m\_\,reverse([a,b,c,d])\. \\
\rho_1:\tt\ m\_\,reverse(Y) \leftarrow m\_\,reverse([X|Y])\. \\
\rho_2:\tt\ reverse([\,], [\,]) \leftarrow m\_\,reverse([\,])\.\\
\rho_3:\tt\ reverse([X|Y],[X|Z]) \leftarrow  m\_\,reverse([X|Y]), reverse(Y,Z)\.
\end{array}
\]
Observe that $P'_{\ref{reverse-example}}$ is not  argument-restricted.
In order to check \acyclicity\ and safety criteria, we have to rewrite rule $\rho_3$ having complex terms in both the head and the body.
Thus we add an additional predicate
$\tt b1$ defined by rule $\rho_4$ and replace $\rho_3$ by $\rho'_3$.
\[
\begin{array}{l}
\rho'_3:\tt\ reverse([X|Y],[X|Z]) \leftarrow  b1(X,Y,Z)\.\\
\rho_4:\tt\ b1(X,Y,Z) \leftarrow  m\_reverse([X|Y]), reverse(Y,Z)\.
\end{array}
\]
The obtained program, denoted $P''_{\ref{reverse-example}}$, is safe but
not \acyclic.
\hfill $\Box$
\end{example}

\begin{example}\label{length-example}
Consider the query $\< \tt length([a,b,c,d],L),$ $ P_{\ref{length-example}}\>$, where $P_{\ref{length-example}}$
is defined by the following rules:
\[
\begin{array}{l}
r_0:\tt\ length([\,],0)\. \\
r_1:\tt\ length([X|T],I+1) \leftarrow length(T,I)\. \ \ \ \ \ \ \ \ \ \ \ \ \ \ \ \ \ \ \ \ \ \\
\end{array}
\]
The equivalent program $P'_{\ref{length-example}}$, is rewritten to be computed by means of a bottom-up evaluator
as follows\footnote{
Observe that program $P'_{\ref{length-example}}$ is equivalent to program 
$P_{\ref{count-ex}}$ presented in the Introduction, assuming that the base predicate $\tt input$ is defined by a fact $\tt input([a,b,c,d])$.
}
:
\[
\begin{array}{l}
\rho_0:\tt\ m\_length([a,b,c,d])\. \\
\rho_1:\tt\ m\_length(T) \leftarrow m\_length([X|T])\. \\
\rho_2:\tt\ length([\,],0) \leftarrow m\_length([\,])\.\\
\rho_3:\tt\ length([X|T],I+1) \leftarrow  m\_length([X|T]),\ length(T,I)\. \\
\end{array}
\]
Also in this case, it is necessary to split rule $\rho_3$ into two rules to avoid
having function symbols in both the head and the body, as shown below:
\[
\begin{array}{l}
\rho'_3:\tt\ length([X|T],I+1) \leftarrow  b1(X,T,I)\.\\
\rho_4:\tt\ b1(X,T,I) \leftarrow  m\_length1(X,T),\ length(T,I)\.
\end{array}
\]
The obtained program, denoted $P''_{\ref{length-example}}$, is safe but
not \acyclic.
\hfill $\Box$
\end{example}

We conclude this section pointing out that
the queries in the two examples above are not recognized as terminating by
most of the previously proposed techniques, including $AR$.
We also observe that many programs follow the structure of programs presented in the examples above.
For instance, programs whose aim is the verification of a given
property on the elements of a given list, have the following structure:
\[
\begin{array}{l}
\tt verify([\,],[\,] )\. \\
\tt verify([X|L_1],[Y|L_2] ) \leftarrow property(X,Y), verify(L_1,L_2)\.
\end{array}
\]
Consequently, queries having a ground argument in the query goal are terminating.

\section{Further Improvements}\label{sec:k-safe}

The safety criterion can be improved further as it is not able
to detect that in the activation graph,
there may be cyclic paths that are not effective or can only be activated a finite number of times.
The next example shows a program which is finitely ground, but recognized as terminating
by the safery criterion.

\begin{example}\label{Example1-Intro}
Consider the following logic program $P_{\ref{Example1-Intro}}$
obtained from $P_{\ref{Example-act-graph}}$ by adding an auxiliary predicate $q$:
\[
\begin{array}{l}
r_1: \tt p(X,X) \leftarrow b(X). \\
r_2: \tt q(f(X),g(X)) \leftarrow p(X,X)\. \\
r_3: \tt p(X,Y) \leftarrow q(X,Y)\.
\end{array}
\]
$P_{\ref{Example1-Intro}}$ is equivalent to
$P_{\ref{Example-act-graph}}$ \wrt predicate $p$.
~\hfill $\Box$
\end{example}

Although the activation graph $\Sigma(P_{\ref{Example1-Intro}})$ contains a cycle, the rules occurring in the cycle cannot
be activated an infinite number of times.
Therefore, in this section we introduce the notion of \emph{active paths}
and extend the definitions of activation graphs and safe programs.

\begin{definition}[Active Path]\label{ActvPath}
Let $\PP$ be a program and  $k \geq 1$ be a natural number. The path
$(r_1,r_2), \dots, (r_k,r_{k+1})$ is an \emph{active path} in the activation graph $\Sigma(\PP)$
iff there is a set of unifiers $\theta_1, \dots \theta_k$, such that
\vspace*{-2mm}
\begin{itemize}
\item
$head(r_1)$ unifies with an atom from $body(r_2)$ with unifier $\theta_1$;
\item
$head(r_i)\theta_{i-1}$ unifies with an atom from $body(r_{i+1})$ with unifier $\theta_i$ for $i\in[2..k]$.
\end{itemize}

\noindent
We write $r_1\rightsquigarrow\!\!\!\!\!\!^{k}\ r_{k+1}$ if there is an active path of length $k$ from $r_1$ to $r_{k+1}$ in $\Sigma(\PP)$.~\hfill $\Box$
\end{definition}

Intuitively, $(r_1,r_2), \dots, (r_k,r_{k+1})$ is an active path
if $r_1$ transitively activates rule $r_{k+1}$, that is
if the head of $r_1$ unifies with some body atom of $r_2$ with mgu $\theta_1$,
then the head of the rule $r_2\theta_1$ unifies with some body atom of $r_3$ with mgu $\theta_2$,
then the head of the rule $r_3\theta_2$ unifies with some body atom of $r_4$ with mgu $\theta_3$,
and so on until
the head of the rule $r_k\theta_{k-1}$ unifies with some body atom of $r_{k+1}$ with mgu~$\theta_k$.

\begin{definition}[k-Restricted Activation Graph]\label{RestrActvGraph}
Let $\PP$ be a program and  $k \geq 1$ be a natural number, the \emph{$k$-restricted activation graph}
$\Sigma_k(\PP)=(\PP,E)$ consists of a set of nodes denoting the rules of $\PP$ and
a set of edges $E$ defined as follows:
there is an edge $(r_i,r_j)$ from $r_i$ to $r_j$ iff $r_i\rightsquigarrow\!\!\!\!\!\!^{k}\ r_j$, i.e. iff there is an active path of length $k$ from $r_i$ to $r_j$.~\hfill $\Box$
\end{definition}

\begin{figure}
  \includegraphics[width=9cm]{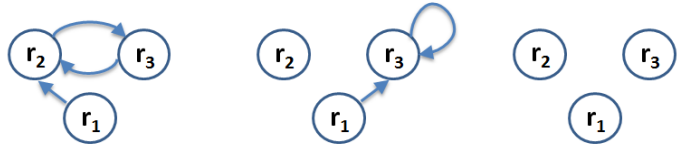}\\
  \caption{$k$-restricted activation graphs: $\Sigma_1(P_{\ref{Example1-Intro}})$ (left), $\Sigma_2(P_{\ref{Example1-Intro}})$ (center),  $\Sigma_3(P_{\ref{Example1-Intro}})$ (right)  }\label{FigureEs13}
\end{figure}

\begin{example}
The $k$-restricted activation graphs for the program of Example \ref{Example1-Intro}, with $k \in [1..3]$, are reported in Figure \ref{FigureEs13}.
\hfill $\Box$
\end{example}

Obviously, the activation graph presented in Definition \ref{actvGraph}
is $1$-restricted.
We next extend the definition of safe function by referring to $k$-restricted activation graphs, instead
of the (1-restricted) activation graph.

\begin{definition}[$k$-Safety Function]\label{k-safe-function}
For any program $\PP\!$ and natural number $k \ge 1$, let $A$ be a set of limited arguments of $\PP\!$.
The \emph{$k$-safety function} $\Psi_{k}(A)$ denotes the set of arguments $q[i] \in args(\PP)$ such that for all rules
$r = q(t_1,\dots,t_m) \leftarrow body(r) \in \PP$,
either $r$ does not depend on a cycle $\pi$ of $\Sigma_j(\PP)$, for some $1 \le j \le k$, or $t_i$ is limited in $r$ w.r.t. $A$.
\hfill $\Box$
\end{definition}

Observe that the \emph{$k$-safety function} $\Psi_{k}$ is defined as a natural extension of the safety function $\Psi$
by considering all the $j$-restricted activation graphs, for $1 \le j \le k$.
Note that the $1$-restricted activation graph
coincides with the standard activation graph and, consequently, $\Psi_1$ coincides with $\Psi$.

\begin{definition}[$k$-Safe Arguments]\label{def-k-safe-arg}
For any program $\PP$, $\mathit{safe}_{k}(\PP) = \Psi_{k}^\infty(\GA(\PP))$
denotes the set of \emph{$k$-safe} arguments of $\PP$.
A program $\PP$ is said to be \emph{$k$-safe} if all arguments are $k$-safe.
\hfill $\Box$
\end{definition}

\begin{example}\label{Example1-Intro-ksafe}
Consider again the logic program $P_{\ref{Example1-Intro}}$ from
Example~\ref{Example1-Intro}. $\Sigma_2(P_{\ref{Example1-Intro}})$
contains the unique cycle $(r_3,r_3)$; consequently, $\tt q[1]$ and
$\tt q[2]$ appearing only in the head of rule $r_2$ are $2$-safe.
By applying iteratively operator $\Psi_2$ to the set of limited arguments $\tt \{ b[1], q[1], q[2] \}$,
we derive that also $\tt p[1]$ and $\tt
p[2]$ are $2$-safe.
Since $\mathit{safe}_2(P_{\ref{Example1-Intro}}) =
args(P_{\ref{Example1-Intro}})$, we have that
$P_{\ref{Example1-Intro}}$ is $2$-safe.
Observe also that
$\Sigma_3(P_{\ref{Example1-Intro}})$ does not contain any edge
and, therefore, all arguments are $3$-safe.
~\hfill $\Box$
\end{example}

For any natural number $k>0$, $\Safe_k$ denotes the class of $k$-safe logic programs, that is the set of programs $\PP$ such that
$safe_k(\PP) = args(\PP)$.
The following proposition states that the classes of $k$-safe programs define a hierarchy
where $\Safe_k \subsetneq \Safe_{k+1}$.

\begin{proposition}\label{Safek-proposition}
The class $\Safe_{k+1}$ of $(k+1)$-safe programs strictly extends the class
$\Safe_{k}$ of $k$-safe programs, for any $k\geq 1$.
\end{proposition}
\vspace{-3mm}
\proof{
($\Safe_k \subseteq \Safe_{k+1}$) It follows straightforwardly from the definition of $k$-safe function.
($\Safe_k \neq \Safe_{k+1}$) To show that the containment is strict, consider the program $P_{\ref{Example1-Intro}}$
from Example~\ref{Example1-Intro} for $k=1$ and the following program $\PP_k$ for $k>1$:
\[
\begin{array}{rl}
r_0\!: & \tt q_1(f(X),g(X)) \leftarrow p(X,X)\.\\
r_1\!: & \tt q_2(X,Y) \leftarrow q_1(X,Y)\.\\
\dots\\
r_{k-1}\!: & \tt q_k(X,Y) \leftarrow q_{k-1}(X,Y)\.\\
\ \ r_{k}\!: & \tt p(X,Y) \leftarrow q_{k}(X,Y)\.\\
\end{array}
\]
It is easy to see that $\PP_k$ is in $\Safe_{k+1}$, but not in $\Safe_k$.
\hfill~$\Box$

\vspace*{3mm}

Recall that the minimal model of a standard program $\PP$ can be characterized in terms of the classical immediate consequence operator $\TT\!_\PP$ defined as follows.
Given a set $I$ of ground atoms, then
$$\begin{array}{ll}
\TT\!_\PP(I)=\{A\theta \mid & \!\!\! \exists r\!: A \leftarrow A_1, \dots, A_n \in \PP \mbox{ and } \exists \theta \mbox{ s.t. }
A_i\theta \in I \mbox{ for every } 1 \leq i \leq n \}
\end{array}$$
where $\theta$ is a substitution replacing variables with constants.
Thus, $\TT\!_\PP$ takes as input a set of ground atoms and returns as output a set of ground atoms; clearly, $\TT\!_\PP$ is monotonic.
The $i$-th iteration of $\TT\!_\PP$ ($i \geq 1$) is defined as follows: $\TT^1_{\!\!\!\PP}(I)=\TT_\PP(I)$ and $\TT_\PP^i(I)=\TT_\PP(\TT\!_\PP^{i-1}(I))$ for $i > 1$.
It is well known that the minimum model of $\PP$ is equal to the fixed point $\TT_\PP^\infty(\emptyset)$.

A rule $r$ is \emph{fired at run-time} with a substitution $\theta$ at step $i$ if 
$head(r)\theta \in T_{\cal P}^i(\emptyset) - T_{\cal P}^{i-1}(\emptyset)$. 
Moreover, we say that \emph{$r$ is fired (at run-time) by a rule $s$}
if $r$ is fired with a substitution $\theta$ at step $i$,
$s$ is fired with a substitution $\sigma$ at step $i-1$,
and $head(s)\sigma \in body(r)\theta$.
Let $\PP$ be a program  whose minimum model is $M = \MM(\PP) = T_{\cal P}^\infty(\emptyset)$,
$M[\![r]\!]$ denotes the set of facts which have been inferred during the application of the
immediate consequence operator using rule $r$, that is the set of facts $head(r)\theta$ such that,
for some natural number $i$, $head(r)\theta \in T_{\cal P}^i(\emptyset) - T_{\cal P}^{i-1}(\emptyset)$;
$M[\![r]\!]$ if infinite iff $r$ is fired an infinite number of times.
Clearly,  if a rule $s$ fires at run-time a rule $r$, then the activation graph
contains an edge $(s,r)$.
An \emph{active sequence} of rules is a sequence of rules $r_1, \dots, r_n$ such that
$r_i$ fires at run-time rule $r_{i+1}$ for $i\in [1..n-1]$.

\begin{theorem}\label{theorem:all_cycles}
Let $\PP$ be a logic program and let $r$ be a rule of $\PP$.
If $M[\![r]\!]$ is infinite, then, for every natural number $k$,
$r$ depends on a cycle of $\Sigma_k(\PP)$.
\end{theorem}
\vspace*{-3mm}
\proof{
Let $n_r$ be the number of rules of $\PP$ and let $N=n_r*k$.
If $M[\![r]\!]$ is infinite we have that there is an active sequence of rules $r'_0, r'_1, \dots, r'_i, \dots, r'_N$
such that $r'_N$ coincides with $r$.
This means that
$$r'_0\rightsquigarrow\!\!\!\!\!\!^{k}\ r'_{k},\
r'_{k} \rightsquigarrow\!\!\!\!\!\!^{k}\ r'_{2k},\
\dots,\
r'_{j*k} \rightsquigarrow\!\!\!\!\!\!^{k}\ r'_{(j+1)*k},\
\dots,\
r'_{(n_r-1)*k} \rightsquigarrow\!\!\!\!\!\!^{k}\ r'_N,$$
\noindent i.e. that the $k$-restricted activation graph $\Sigma_k(\PP)$ contains path
$\pi=(r'_0,r'_{k}),$ $(r'_{k},r'_{2k}),\dots,
(r'_{j*k},r'_{(j+1)*k}),\dots, (r'_{(n_r-1)*k},r)$.
Observe that the number of rules involved in $\pi$ is $n_r+1$ and is greater than
the number of rules of $\PP$.
Consequently, there is a rule occurring more than once in $\pi$,  i.e. $\pi$ contains a cycle.
Therefore, $r$ depends on a cycle  of $\Sigma_k(\PP)$.\hfill $\Box$
}

\vspace*{2mm}

As shown in Example~\ref{Example1-Intro-ksafe}, in some cases the analysis of the $k$-restricted activation graph
is enough to determine the termination of a program.
Indeed, let $cyclicR(\Sigma_k(\PP))$ be the set of rules $r$ in $\PP$ s.t. $r$ depends on a cycle in $\Sigma_k(\PP)$, the following results hold.

\begin{corollary}
A program $\PP$ is terminating if $\forall r \in \PP$,
$\exists k$ s.t. $r \not\in cyclicR(\Sigma_k(\PP))$.
\end{corollary}
\vspace{-3mm}
\proof{Straightforward from Theorem \ref{theorem:all_cycles}.
\hfill $\Box$
}

\vspace*{3mm}
Obviously, if there is a $k$ such that for all rules $r \in \PP$
$r \not\in cyclicR(\Sigma_k(\PP))$, $\PP$ is terminating.
We conclude this section showing that the improvements here discussed
increase the complexity of the technique which is not polynomial anymore.

\begin{proposition}
For any program $\PP$ and natural number $k>1$, the activation graph $\Sigma_k(\PP)$ can be
constructed in time exponential in the size of $\PP$ and $k$.
\end{proposition}
\vspace{-3mm}
\proof{
Let $(r_1,r_2) \cdots (r_k,r_{k+1})$ be an active path of length $k$ in $\Sigma(\PP)$.
Consider a pair $(r_i,r_{i+1})$ and two unifying atoms $A_i=head(r_i)$ and $B_{i+1}\in body(r_{i+1})$
(with $1\leq i\leq k$), the size of an mgu $\theta$ for $A_i$ and $B_{i+1}$,
represented in the standard way (cif. Section \ref{sec-preliminaries}), can be exponential
in the size of the two atoms.
Clearly, the size of $A_i\theta$ and $B_{i+1}\theta$ can also be exponential.
Consequently, the size of $A_{i+1}\theta$ which is used for the next step, can grow exponentially as well.
Moreover, since in the computation of an active path of length $k$ we apply $k$ mgu's,
the size of terms can grow exponentially with $k$.
\hfill $\Box$
}

\vspace*{3mm}
Observe that for the computation of the 1-restricted argument graph it is sufficient to determine
if two atoms unify (without computing the mgu), whereas for the computation of the $k$-restricted argument graphs, with $k>1$,
it is necessary to construct all the mgu's and to apply them to atoms.

\section{Computing stable models for disjunctive programs}\label{sec:compSM}

In this section we discuss how termination criteria, defined for standard programs,
can be applied to general disjunctive logic programs.
First, observe that we have assumed that whenever the same variable $X$ appears in two terms occurring, respectively,
in the head and body of a rule, at most one of the two terms is a complex term and that the nesting level of complex terms is at most one.
There is no real restriction in such an assumption as every program could be rewritten into an equivalent program satisfying such a condition.
For instance, a rule $r'$ of the form
$$
\tt p(f(g(X)),h(Y,Z)) \leftarrow p(f(X),Y),\ q(h(g(X),l(Z)))
$$
is rewritten into the set of 'flatten' rules below:
\[
\begin{array}{lll}
\tt p(f(A),h(Y,Z)) & \leftarrow & \tt b_1(A,Y,Z) \\
\tt b_1(g(X),Y,Z)  & \leftarrow & \tt b_2(X,Y,Z) \\
\tt b_2(X,Y,Z)     & \leftarrow & \tt b_3(X,Y,g(X),l(Z)) \\
\tt b_3 (X,Y,B,C)  & \leftarrow & \tt p(f(X),Y),\ q(h(B,C))
\end{array}
\]
where $\tt b_1,b_2$ and $\tt b_3$ are new predicate symbols, whereas
$\tt A, B$ and $\tt C$ are new variables introduced to flat terms with depth greater than 1.

More specifically, let $d(p(t_1,\dots,t_n)) = max \{ d(t_1),\dots, d(t_n) \}$ be the depth of atom $p(t_1,\dots,t_n)$
and $d(A_1,\dots,A_n) = max \{ d(A_1),\dots, d(A_n) \}$ be the depth of a conjunction of atoms $A_1,\dots,A_n$,
for each standard rule $r$ we generate a set of 'flatten' rules, denoted by $flat(r)$
whose cardinality is bounded by $O(d(head(r)) + d(body(r))$.\\
Therefore, given a standard program $\PP$, the number of rules of the rewritten program is
polynomial in the size of $\PP$ and bounded by
$$O \left(\ \sum_{r \in {\cal P}} \ d(head(r)) + d(body(r))  \right).$$

Concerning the number of arguments in the rewritten program,
for a given rule $r$ we denote with $nl(r,h,i)$ (resp. $nl(r,b,i)$) the number of occurrences of function symbols occurring at the same
nesting level $i$ in the head (resp. body) of $r$ and with
$nf(r) = max\{ nl(r,t,i)\ |\ t \in \{h,b\} \wedge i > 1\}$.
For instance, considering the above rule $r'$, we have that
$nl(r',h,1) = 2$ (function symbols $\tt f$ and $\tt h$ occur at nesting level $1$ in the head),
$nl(r',h,2) = 1$ (function symbol $\tt g$ occurs at nesting level $2$ in the head),
$nl(r',b,1) = 2$ (function symbols $\tt f$ and $\tt h$ occur at nesting level $1$ in the head),
$nl(r',b,2) = 2$ (function symbols $\tt g$ and $\tt l$ occur at nesting level $2$ in the head).
Consequently, $nf(r') = 2$.

The rewriting of the source program results in a 'flattened' program with
$|flat(r)|-1$ new predicate symbols.
The arity of every new predicate in $flat(r)$ is bounded by $|var(r)| + nf(r)$.
Therefore, the global number of arguments in the flattened program is bounded by
$$O \left(\ args(\PP) + \sum_{r \in {\cal P}} \left(\ |var(r)| + nf(r)\ \right) \right).$$

\vspace*{3mm}
The termination of a disjunctive program $\PP$ with negative literals
can be determined by rewriting it into a standard
logic program $st(\PP)$ such that every stable model of $\PP$ is
contained in the (unique) minimum model of $st(\PP)$, and then by checking
$st(\PP)$ for termination.

\begin{definition}[Standard version]
Given a program $\PP$, $st(\PP)$ denotes the standard
program, called \emph{standard version}, obtained by replacing
every disjunctive rule $r = a_1 \vee \cdots \vee a_m \leftarrow body(r)$
with $m$ standard rules of the form $a_i \leftarrow body^{+}(r)$,
for $1 \leq i \leq m$.\\
Moreover, we denote with $ST(\PP)$ the program derived from $st(\PP)$
by replacing every derived predicate symbol $q$ with a new derived predicate symbol~$Q$.
~\hfill $\Box$
\end{definition}

The number of rules in the standard program $st(\PP)$ is equal to $\sum_{r \in {\cal P}} |head(r)|$,
where $|head(r)|$ denotes the number of atoms in the head of $r$.

\begin{example}\label{standard-rule-example}
Consider program $P_{\ref{standard-rule-example}}$ consisting
of the two rules
\[
\begin{array}{l}
\tt p(X) \vee q(X) \leftarrow r(X), \neg a(X). \\
\tt r(X) \leftarrow b(X), \neg q(X).
\end{array}
\]
where $\tt p$, $\tt q$ and $\tt r$ are derived (mutually recursive) predicates,
whereas $\tt a$ and $\tt b$ are base predicates. The derived standard
program $st(P_{\ref{standard-rule-example}})$ is as follows:
\[
\begin{array}{l}
\tt p(X) \leftarrow r(X). \\
\tt q(X) \leftarrow r(X). \\
\tt r(X) \leftarrow b(X). \vspace*{-6mm}
\end{array}
\]
\hfill $\Box$
\end{example}

\begin{lemma}\label{lemma-stable}
For every program $\PP$, every stable model $M \in \SM(\PP)$ is contained in
the minimum model $\MM(st(\PP))$.
 \end{lemma}
\vspace*{-4mm}
\proof{
From the definition of stable models we have that every $M \in \SM(\PP)$
is the minimal model of the ground positive program $\PP^M$.
Consider now the standard program $\PP'$ derived from $\PP^M$ by replacing every
ground disjunctive rule $r = a_1 \vee \dots \vee a_n \leftarrow body(r)$ with $m$ ground normal rules
$a_i \leftarrow body(r)$.
Clearly,  $M \subseteq \MM(\PP')$.
Moreover, since $\PP' \subseteq st(\PP)$, we have that $\MM(\PP') \subseteq \MM(st(\PP))$.
Therefore, $M \subseteq \MM(st(\PP))$.~\hfill $\Box$
}

\vspace*{2mm}
The above lemma implies that for any logic program $\PP$, if
$st(\PP)$ is finitely ground we can restrict the Herbrand base and only consider head (ground) atoms $q(t)$ such that
$q(t) \in \MM(st(\PP))$.
This means that, after having computed the minimum model of $st(\PP)$,
we can derive a finite ground instantiation of $\PP$,
equivalent to the original program, by considering only ground atoms contained in $\MM(st(\PP))$.

We next show how the original program $\PP$ can be rewritten so that, after having computed $\MM(st(\PP))$, every
grounder tool easily generates an equivalent finitely ground program.
The idea consists in generating, for any disjunctive program $\PP$ such that
$st(\PP)$ satisfies some termination criterion (e.g. safety),
a new equivalent program $ext(\PP)$.
The computation of the stable models of $ext(\PP)$ can be carried out by considering
the finite ground instantiation of $ext(\PP)$ \cite{dlv,smodels,clasp}.

For any disjunctive rule $r = q_1(u_1) \vee \cdots \vee q_k(u_k) \leftarrow body(r)$, the conjunction of atoms
$Q_1(u_1),\.\.\.,Q_k(u_k)$
will be denoted by $headconj(r)$.

\begin{definition}[Extended program]\label{def:extendend_program}
Let $\PP$ be a disjunctive program and let $r$ be a rule of $\PP$,
then, $ext(r)$ denotes the (disjunctive) \emph{extended rule} $head(r)
\leftarrow headconj(r), body(r)$ obtained by extending the body of
$r$, whereas $ext(\PP) = \{ ext(r)\ |\ r \in \PP \} \cup ST(\PP)$
denotes the (disjunctive) \emph{extended program} obtained by extending the rules of
$\PP$ and adding (standard) rules defining the new predicates.
\hfill $\Box$
\end{definition}

\begin{example}
Consider the program $P_{\ref{standard-rule-example}}$ of Example
\ref{standard-rule-example}. The extended program
$ext(P_{\ref{standard-rule-example}})$ is as follows:
\[
\begin{array}{l}
\tt p(X) \vee q(X) \leftarrow P(X), Q(X), r(X), \neg a(X). \\
\tt r(X) \leftarrow R(X), b(X), \neg q(X). \\
\tt P(X) \leftarrow R(X). \\
\tt Q(X) \leftarrow R(X). \\
\tt R(X) \leftarrow b(X). \vspace*{-6mm}
\end{array}
\]
\hfill $\Box$
\end{example}

The following theorem states that $\PP$ and $ext(\PP)$ are
equivalent w.r.t. the set of predicate symbols in $\PP$.

\begin{theorem}\label{Th-stable-equiv}
For every program $\PP$, $\SM(\PP)[S_{\cal P}] =
\SM(ext(\PP))[S_{\cal P}]$, where $S_{\cal P}$ is the set of
predicate symbols occurring in $\PP$. 
\end{theorem}
\proof{
First, we recall that $ST(\PP) \subseteq ext(\PP)$ and  assume that
$N$ is the minimum model of $ST(\PP)$, i.e. $N = \MM(ST(\PP))$.
\begin{itemize}
\item
We first show that for each $S \in \SM(ext(\PP))$, $M = S - N$ is a stable model for $\PP$,
that is $M \in \SM(\PP)$.\\
Let us consider the ground program $\PP^{\prime\prime}$ obtained from $ext(\PP)^S$ by first deleting
every ground rule  $r = head(r)\leftarrow headconj(r), body(r)$ such that
$N \not\models headconj(r)$ and then by removing from the remaining rules, the conjunction $headconj(r)$.
Observe that the sets of minimal models for $ext(\PP)^S$ and $\PP^{\prime\prime}$
coincide, i.e. $\MM(ext(\PP)^S) = \MM(\PP^{\prime\prime})$.
Indeed, for every $r$ in $ext(\PP)^S$, if $N \not\models headconj(r)$, then
the body of $r$ is false and thus $r$ can be removed as it does not contribute to infer head atoms.
On the other hand, if $N \models headconj(r)$, the conjunction $headconj(r)$ is trivially true, and can be safely deleted from the body of $r$.

Therefore, $M \cup N \in \MM(\PP^{\prime\prime})$.
Moreover, since $\PP^{\prime\prime} = (\PP \cup ST(\PP))^S = \PP^M \cup ST(\PP)^N$, we have that $M \in \MM(\PP^M)$,
that is $M \in \SM(\PP)$.
\item
We now show that for each $M \in \SM(\PP)$, $(M \cup N) \in \SM(ext(\PP))$.\\
Let us assume that $S = M \cup N$.
Since $M \in \MM(\PP^M)$ we have that $S \in \SM(\PP \cup ST(\PP))$,
that is $S \in \MM((\PP \cup ST(\PP))^S)$.
Consider the ground program $\PP'$ derived from $(\PP \cup ST(\PP))^S$ by replacing every rule
disjunctive $r = head(r)\leftarrow body(r)$ such that $M \models body(r)$
with $ext(r) = head(r) \leftarrow headconj(r), body(r)$.
Also in this case we have that $\MM(\PP \cup ST(\PP))^S) = \MM(\PP')$
as $S \models body(r)$ iff $S \models body(ext(r))$.
This, means that $S$ is a stable model for $ext(\PP)$.
\hfill $\Box$
\end{itemize}
}

\section{Conclusion}\label{sec-conclusions}

In this paper we have proposed a new approach for checking, on the
basis of structural properties, termination of the bottom-up evaluation of logic programs with function symbols. We have first proposed a technique, called
\emph{\acyclicity}, extending the class of argument-restricted
programs by analyzing the propagation of complex terms among
arguments using an extended version of the argument graph. Next, we
have proposed a further extension, called \emph{safety}, which also
analyzes how rules can activate each other (using the activation
graph) and how the presence of some arguments in a rule limits its
activation. We have also studied the application of the techniques
to partially ground queries and have proposed further
improvements which generalize the safety criterion through the
introduction of a hierarchy of classes of terminating programs,
called \emph{$k$-safety}, where each $k$-safe class strictly includes
the $(k\-1)$-safe class.

Although our results have been defined for standard programs,
we have shown that they can also be applied to disjunctive programs
with negative literals, by simply rewriting the source programs.
The semantics of the rewritten program is \"equivalent" to the
semantics of the source one and can be computed by current answer
set systems. Even though our framework refers to the model theoretic
semantics, we believe that the results presented here go beyond the
ASP community and could be of interest also for the (tabled) logic
programming community (e.g. tabled Prolog community).

\paragraph{\bf Acknowledgements.}
The authors would like to thank the anonymous reviewers
 for their valuable comments and
suggestions.

\end{document}